\documentclass[twocolumn]{aastex63}
\let\splitbox=\relax
\usepackage[export]{adjustbox}

\usepackage[utf8]{inputenc}
\usepackage{float}
\usepackage[caption=false]{subfig}
\usepackage{multirow}
\usepackage{amssymb}

\newcommand{\kms} {\,km\,s$^{-1}$}
\newcommand{\masyr} {\,mas\,yr$^{-1}$}

\begin{document}
\title{Enhancement of Double-Close-Binary Quadruples}

\author[0000-0002-1716-9430]{Gavin B. Fezenko}
\affiliation{Department of Physics \& Astronomy, Johns Hopkins University, Baltimore, MD 21218, USA}

\author[0000-0003-4250-4437]{Hsiang-Chih Hwang}
\affiliation{Department of Physics \& Astronomy, Johns Hopkins University, Baltimore, MD 21218, USA}
\affiliation{Institute for Advanced Study, Princeton, 1 Einstein Drive, NJ 08540, USA}

\author[0000-0001-6100-6869]{Nadia L. Zakamska}
\affiliation{Department of Physics \& Astronomy, Johns Hopkins University, Baltimore, MD 21218, USA}
\affiliation{Institute for Advanced Study, Princeton, 1 Einstein Drive, NJ 08540, USA}

\begin{abstract}
    Double-close-binary quadruples (2+2 systems) are hierarchical systems of four stars where two short-period binary systems move around their common center of mass on a wider orbit. Using {\it Gaia} Early Data Release 3, we search for comoving pairs where both components are eclipsing binaries. We present eight 2+2 quadruple systems with inner orbital periods of $<0.4$ days and with outer separations of $\gtrsim 1000$ AU. All but one system are newly discovered by this work, and we catalog their orbital information measured from their light curves. We find that the occurrence rate of 2+2 quadruples is 7.3$\pm$2.6 times higher than what is expected from random pairings of field stars. At most a factor of $\sim$2 enhancement may be explained by the age and metallicity dependence of the eclipsing binary fraction in the field stellar population. The remaining factor of $\sim$3 represents a genuine enhancement of the production of short-period binaries in wide-separation ($>10^3$ AU) pairs, suggesting a close binary formation channel that may be enhanced by the presence of wide companions.

\end{abstract}

\section{Introduction} 
\label{sec:intro}

    High-order star systems constitute a large portion of the stellar population, with triple and quadruple systems making up somewhere around 13\% of systems and quadruples specifically accounting for about 4\% \citep{Tokovinin2014}. Systems with more than two stars tend to arrange themselves in hierarchical configurations for long-term dynamical stability. This occurs because non-hierarchical structures with more than two stars at similar distances from each other are dynamically unstable, and will either decay into binaries and single stars \citep{Fujii2011} or rearrange themselves into hierarchical structures \citep{Harrington1972}. Therefore, with the exception of some very young systems, we typically do not observe stellar triples and quadruples with comparable separations between stars. 
    
    Two configurations offering long-term stability are possible for stellar quadruples \citep{Chambliss1992}. In 3+1 systems, an `inner' close binary is orbited at a much larger distance by another body, and then the resulting hierarchical triple is orbited by yet another star on an even wider orbit \citep{Tokovinin2014, Hamers2018}. Double-close-binary quadruples, also called 2+2 systems, are hierarchical systems composed of two close inner binary pairs gravitationally bound to one another and orbiting the barycenter of the two pairs on a much wider `outer' orbit. A few dozen 2+2 quadruples are known. Some are identified because the two inner binaries are separated by wide enough distances that the binaries are spatially resolved from one another and appear to be co-moving on the sky, their mutual orbital velocity being much smaller than their Galactic motion \citep{Poveda1994, Lepine2007}. For others, their projected outer separation is within the spatial resolution element of a typical optical observation, so they are identified either by their radial velocity signatures \citep{Abt1976, Lu2001} or by separating the superposition of two eclipsing binary light curves within one spatial resolution element \citep{Zasche2019}.
    
    Some interesting and not yet well understood star formation effects and astrophysical dynamics come into play in 2+2 quadruples. They appear to be over-represented in a volume-limited sample by \citet{Tokovinin2014}, suggesting some kind of preferential formation mechanism. Quadruples are more common than what would be expected from the capture of a binary by another binary during the dissolution of the natal cluster \citep{Kroupa1995}. Some special formation mechanisms are also suggested by the inconsistency between the observed inner periods and those expected from the long-term dynamical evolution \citep{Hamers2018}. The 2+2 quadruples that have relatively small `outer' separations may be affected by strong dynamical interactions \citep{Pejcha2013} and can get stuck into 1:1 and 3:2 resonances \citep{Tremaine2020}. Finally, 2+2 quadruples of extremely massive stars have emerged as promising progenitors of the most massive known gravitational wave mergers \citep{Fragione2019}.  

    In this paper we present a new sample of hierarchical 2+2 quadruples with outer separations $\gtrsim$1000 AU selected from {\it Gaia} data and further investigate their light curves from variability surveys. We describe sample selection and light curve analysis in Section \ref{sec:data}. We present our new samples and their statistical analysis in Section \ref{sec:results}, where we find that 2+2 quadruples are significantly more common than would be expected from random pairings of field binaries. We discuss possible origins and the implications of this enhancement in Section~\ref{sec:discussion} and conclude in Section~\ref{sec:conclusions}. 
    
\section{Sample selection}
\label{sec:data}

\subsection{Gaia parent sample selection}
\label{sec:gaia}

    The initial sample for this investigation is selected from the {\it Gaia} \citep{Gaia2016,Gaia2018,Gaia2021} Early Data Release 3. The full initial Astronomical Data Query Language (ADQL) query used can be found in Appendix \ref{query}. The parent sample is limited to within 1 kpc by requiring  \texttt{parallax} $>$1 mas, with  \texttt{parallax\_over\_error $>$ 5}, \texttt{ruwe $<$ 1.4}, and \texttt{visibility\_periods\_used $\geq 11$} to ensure robust astrometric solutions. For reliable photometry, we require sufficient signal-to-noise ratios in {\it G}, {\it BP}, and {\it RP} bands (\texttt{phot\_g\_mean\_flux\_over\_error} $>50$, \texttt{phot\_bp\_mean\_flux\_over\_error} $>20$, \texttt{phot\_rp\_mean\_flux\_over\_error} $>20$). We limit \texttt{phot\_bp\_rp\_excess\_factor} $<$ 1.4 so that the {\it BP} and {\it RP} photometry is not significantly affected by nearby sources \citep{Riello2021}. 
    
    In order to select main-sequence stars, we limit our sample to the color range of \textit{BP$-$RP}$=$0.7 to 1.6\,mag.  Because we will use root-mean-squared variability to select short-period eclipsing binaries, this color range is important to ensure that the high-amplitude variables are mainly short-period eclipsing binaries, excluding pulsating stars on the bluer side and flaring stars on the redder side 
    \citep{Gaia2019}. Additionally, to avoid comoving groups and stellar clusters, we require that \texttt{N\_comoving\_group < 50}, a parameter measured in the comoving search stage detailed in Sec.~\ref{sec:comoving}. $\Delta G$ described by \citet{Hamer2019} is the offset between the observed absolute $G$-band magnitude of the source and the $G$-band magnitude of the Pleiades at the same \textit{BP$-$RP} color. We require $\mid \Delta G \mid < 1.5$ mag to select stars close to main sequence but allows for binaries (equal-mass binary is brighter than a single star of the same color by 0.75 mag) and stars with different metallicities. This cut avoids any evolved stars, such as white dwarfs and giants. Under these requirements the parent sample contains 15,684,999 main-sequence sources. 

\subsection{Comoving pairs}
\label{sec:comoving}

    Using the astrometric data from {\it Gaia}, we identify resolved comoving pairs of stars following a procedure similar to that described in \citet{Hwang2020wide}. For each target star in the query, its nearby stars are selected from the parent sample (and thus they satisfy the same selection criteria described in Sec.~\ref{sec:gaia}) where the parallax difference is $<0.2$ mas or where the parallaxes are within $6\sigma$. We adopt these parallax difference criteria because parallax errors in Gaia EDR3 are underestimated for pairs separated by $<4$\,arcsec, and astrometric measurements may be affected by the presence of subsystems \citep{El-Badry2021}. A target star with one nearby star is considered as a pair. For each pair, we compute its physical separation and physical relative velocity projected on the plane of the sky. Specifically, the (projected) physical separation is the angular separation between the two stars divided by their mean parallax, and the (projected) physical relative velocity is the proper motion vector difference divided by their mean parallax. In these calculations, we compute distances using the inverse of parallaxes. Because we only use stars with $>5\sigma$ parallax detections, we do not find significant changes in our results if we use distances from \citet{Bailer-Jones2018}.

    To remove contamination of comoving pairs in comoving groups and stellar clusters, we compute the parameter \texttt{N\_comoving\_group}, originally used in \citet{Hwang2020wide}, for each star. This parameter is the number of stars in the parent sample which are, with respect to the target star: (1) between $10^5$ AU and $10^6$ AU away, and (2) have relative velocity $<10$ km/s. A large value of \texttt{N\_comoving\_group} would suggest the star is among a large group or a cluster of stars. Limiting this parameter eliminates pairs that are moving along with clusters, which would contaminate the comoving sample. A value of \texttt{N\_comoving\_group $<50$} is adopted for our analysis. 
    
\begin{figure*}[t!]
    \centering
            \includegraphics[width = 0.48\textwidth]{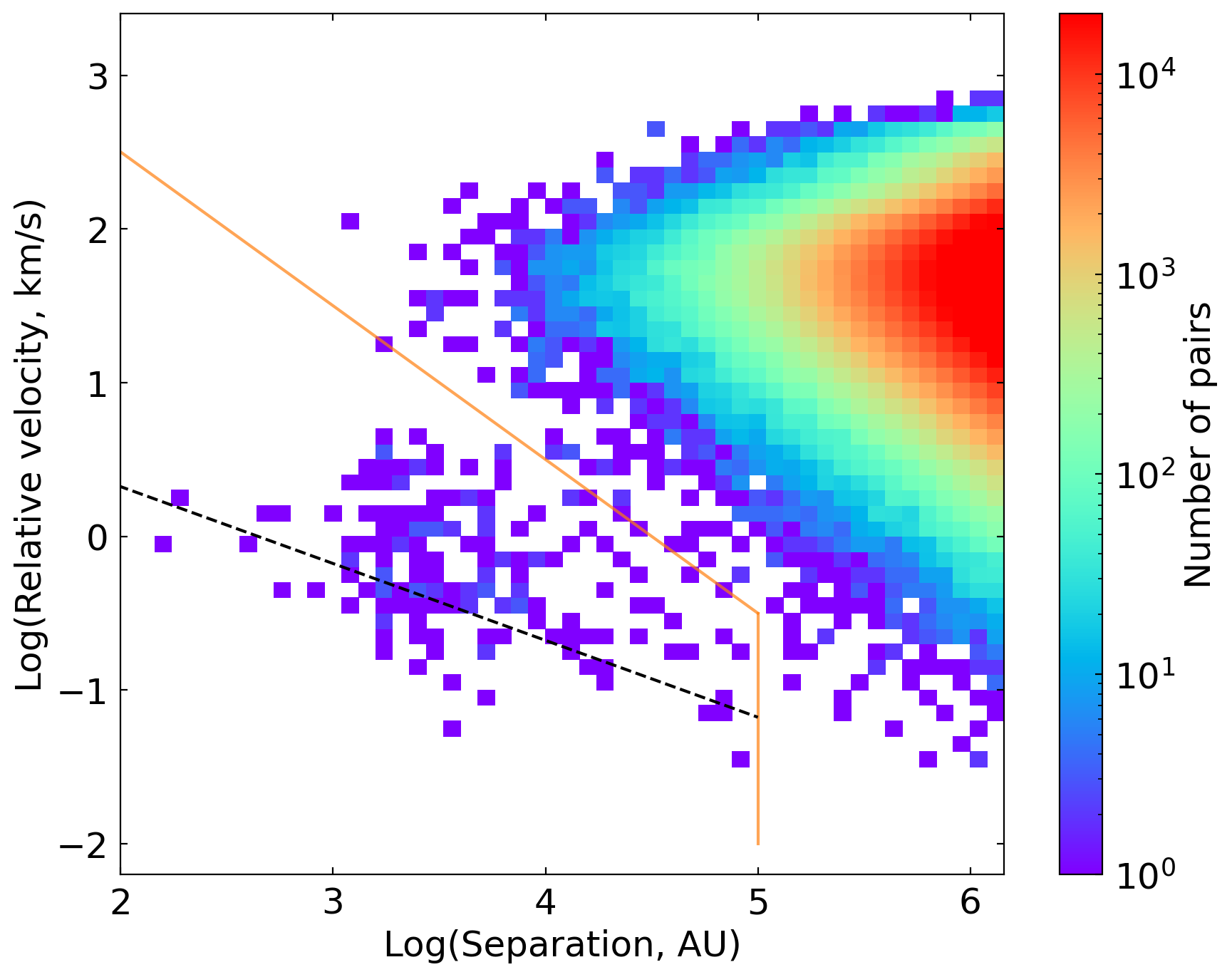}
            \includegraphics[width = 0.48\textwidth]{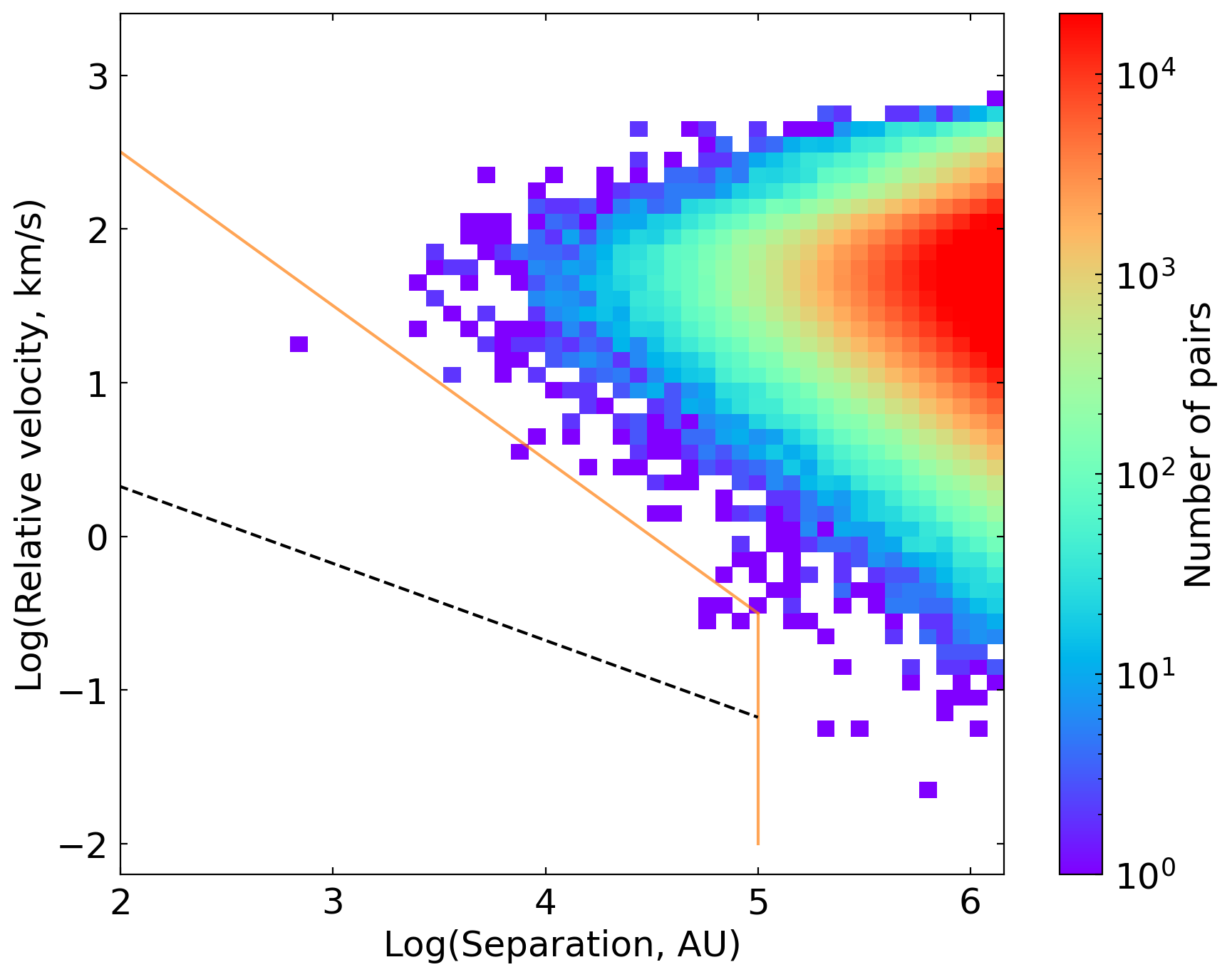}
        \caption{Left: The comoving search for 10,000 random stars. Pairs below the solid orange demarcation line are consistent with being comoving pairs as per our criterion. The dashed line represents the expected Keplerian relative velocity. Right: same as left, but after shifting all target stars by 1 degree North and performing a new comoving search, so that any resulting pairs should be chance alignments. The rarity of pairs below the demarcation line in the right panel indicates that the pairs identified in the original search are high-confidence wide binaries. \label{fig:shift}}
    \end{figure*}

    We select 10,000 random stars and conduct a comoving search against the parent sample. The resulting separations and relative velocities of all the pairs are shown in Figure \ref{fig:shift}, left. The solid line is the demarcation line proposed by \cite{Hwang2020wide} to select comoving pairs. Most of the pairs are chance-alignment pairs located above the demarcation line with high relative velocities and separations. We find 221 pairs located below the demarcation line that are high-confidence comoving wide binaries.
    
    We test how chance-alignment pairs contaminate the comoving search by shifting the coordinates of these 10,000 stars. We offset each target star by 1 degree to the North along declination and repeat the comoving search on the shifted targets, leaving the parent sample at its original true location. Therefore, all pairs found from this test are chance-alignment pairs. The results of this search are shown in Figure \ref{fig:shift}, right, reproducing the overall structure of the chance-alignment pairs in the left panel very well. We only find 7 pairs falling under the demarcation line, implying that the contamination level is about $7/221=3.2$ per cent among the resulting wide binary sample. Therefore, contamination from chance-alignment plays a minor role in our comoving search results.
    
    Our comoving search method is slightly different from \cite{El-Badry2021}. They identify wide binaries that have relative velocities consistent with Keplerian motions, whereas we select wide binaries that are unlikely to be chance alignments. In Figure \ref{fig:shift}, the black dashed line represents the expected Keplerian velocity for an equal-solar-mass binary, with an additional factor of $\pi/4$ coming from averaging over random binary inclinations. While many pairs seem to align with the dashed line very well, some pairs are well above the dashed line but still below the demarcation line. Binaries and higher-order multiples may affect the astrometric measurements through various reasons, either through their orbital motion \citep{Belokurov2020} or photometric variability \citep{Hwang2020var}. In addition, since about half of wide binaries have close companions \citep{Moe2021}, the deviation from the point spread function due to the partially resolved close companions may downgrade the astrometric measurements. Therefore, binary and higher-order multiple systems may have a measured velocity higher than the expected Keplerian value due to these various reasons, and would be excluded by \cite{El-Badry2021}. Thus, to be more complete to binaries with these potential issues, we conduct our own comoving search with relaxed relative velocity selection criteria. The contamination test in Section \ref{sec:comoving} suggests that the contamination rate is only 3.2 percent using this relaxed relative velocity selection. We compare our results with those based on the more strict velocity selection from \cite{El-Badry2021} in Sec.~\ref{sec:elbadry}, and the main conclusions remain the same. 
    
\begin{figure}[h!]
        \includegraphics[width=0.5\textwidth]{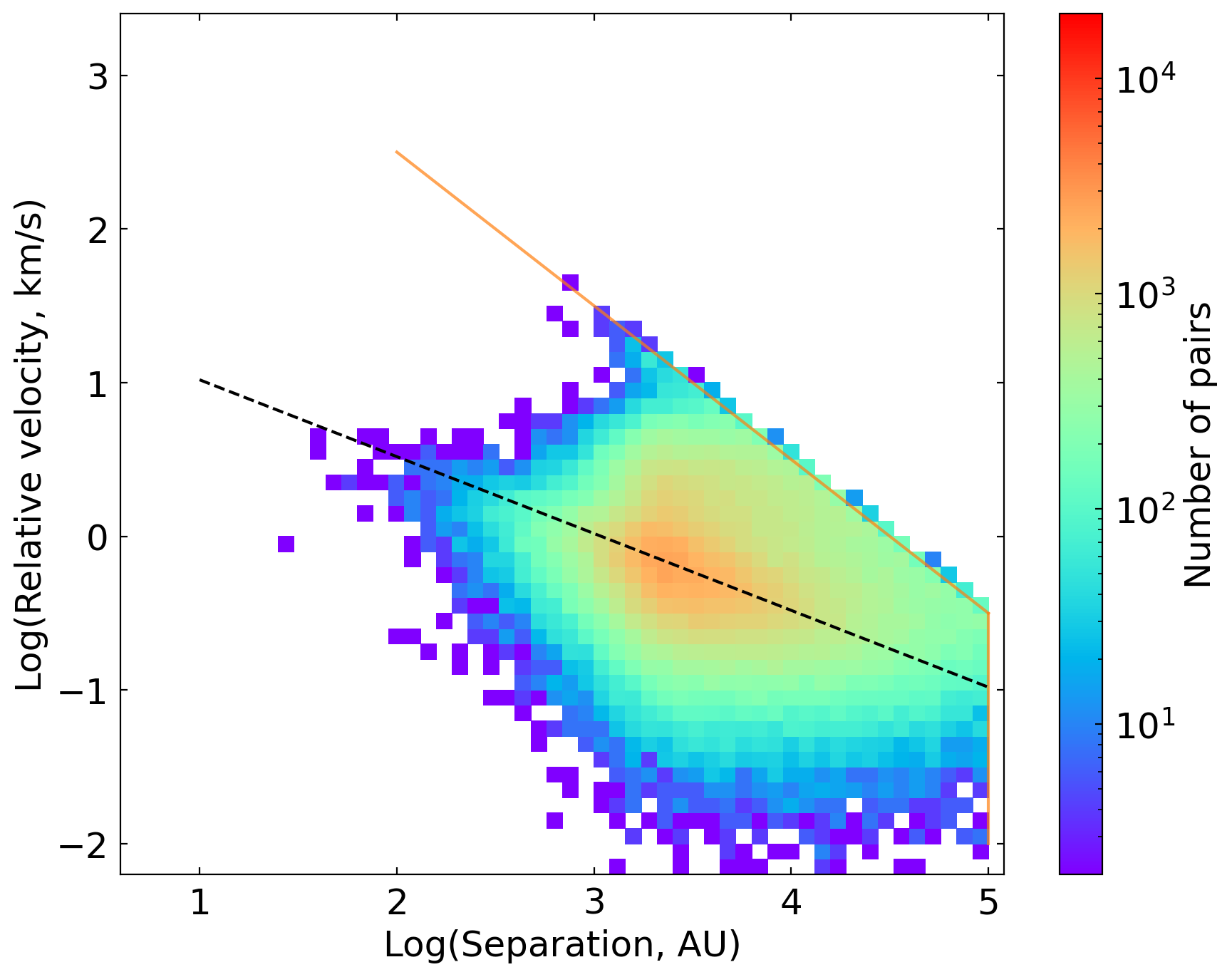}
        \centering
        \caption{Separations and relative velocities for co-moving wide binaries. The solid orange line represents the selection demarcation line. Pairs outside the demarcation line are mainly chance-alignment pairs and are not plotted here. The dashed black line represents the Keplerian relative velocity at a given separation. Overall, wide binaries follow the Keplerian velocities very well. The enhanced number of pairs at $\sim2000$\,AU is due to the spatial resolution determined by {\it Gaia}'s {\it BP/RP} photometry.  }   
        \label{sepvel}
    \end{figure}

    Figure \ref{sepvel} shows our full comoving search for the entire parent sample, resulting in a total of 131,245 pairs of unique wide binaries (thus $N_{wide}=2\times131,245=262,490$ stars in wide binaries). As the search is limited by the computational memory, we only keep the comoving pairs below the demarcation line in the subsequent analysis. We exclude sources that have more than two comoving companions, e.g. resolved triples. These systems are rare ($\sim1$\%) compared to the number of wide binaries, and none of them contains EBs with variability $>5$\%, so excluding them does not affect our results. Overall, the relative velocities of these comoving pairs follow the expected Keplerian motion. Some chance-alignment pairs are still present close to the demarcation line, especially at separations of $\sim10^3$\,AU, but their number is far smaller than the entire wide binary sample and does not affect the main results. The enhanced number of pairs at $\sim2000$\,AU is mainly due to {\it Gaia}'s spatial resolution limit. {\it Gaia} EDR3 measures {\it BP} and {\it RP} magnitudes by summing the total flux in a $\sim2$ arcsec window, without any deblending treatment. Because we use \texttt{phot\_bp\_rp\_excess\_factor} $<$ 1.4 to ensure reliable \textit{BP-RP} colors, this criterion also excludes stars that have nearby sources within $\sim2$\,arcsec. Since our sample is within 1\,kpc, 2\,arcsec corresponds to 2000\,AU, resulting in the enhanced number of pairs at 2000\,AU in Figure \ref{sepvel}. 
    
\subsection{Eclipsing binary selection}
\label{sec:binaries}

    We select eclipsing binaries using {\it Gaia}'s photometric variability information. Due to the {\it Gaia} telescope's six-hour revolution and its two fields of view \citep{Gaia2016}, {\it Gaia}'s cadence is sensitive to photometric variability on timescales of a few hours, ideal for main-sequence eclipsing binaries. {\it Gaia} EDR3 has not released the full light curves and the eclipsing binary catalog, but the root-mean-squared variability of the $G$-band flux can be computed from $\sigma_G$:
    \begin{equation}
        \sigma_G = \texttt{phot\_g\_mean\_flux\_error} \times \sqrt{\texttt{phot\_g\_n\_obs}}.
    \end{equation}
    The variability selection for this study follows that detailed in \citet{Hwang2020short}, Section 2.4. We define the G-band fractional variability as $f_{G,raw} = \sigma_G/F_G$ where $F_G$ is the mean $G$-band flux reported from {\it Gaia}. We measure the instrumental variability $f_{G,inst}$, which is a function of $G$-band magnitude, by measuring the median $f_{G,raw}$ for a large number of field stars with different $G$-band magnitudes. In the end, we remove the instrumental variability for each star by computing $f_{G}=\sqrt{f_{G,raw}^2-f_{G,inst}^2}$, where $f_{G,inst}^2$ is evaluated at the star's $G$-band magnitude. For sources that have $f_{G,raw}<f_{G,inst}$, we set their $f_G$ to zero because they do not have significant variability compared to the instrumental level. The faintest star in our parent sample is $G=$18.7\,mag, where $f_{G,inst}=0.034$, meaning that the instrument variability contributes to a $\sim3$ per cent level and we are able to probe variability larger than 3 per cent. In this work, we select an EB sample that has variability $f_{G}>0.05$, which results in 26,734 EBs. 
    
    \begin{figure}[h!]
        \includegraphics[width=0.48\textwidth]{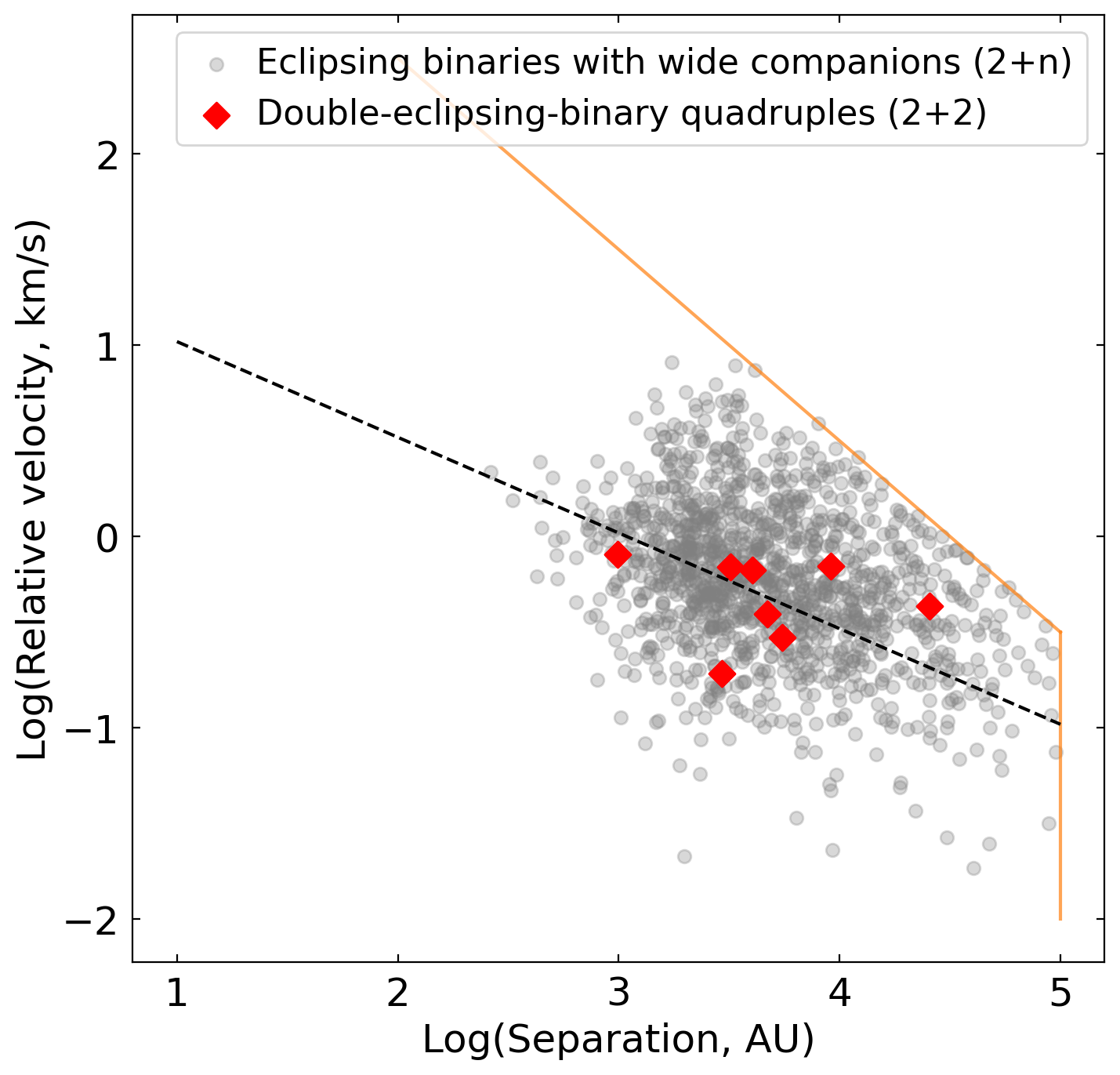}
        \centering
        \caption{Separations and relative velocities of EBs with one comoving companion. Plotted in gray are the pairs where at least one source is an EB comoving with any companion (2+n systems). Red diamonds are 8 double-EB pairs (2+2 systems), and 7 of them are newly discovered. }\label{fig:sepvel2}
    \end{figure}

    Using fractional variability to select eclipsing binaries on the main sequence assumes that eclipsing binaries are the dominant high-amplitude variability sources on the main sequence. While this assumption is generally true \citep{Hwang2020short}, there can be some contamination from other variables due to the lack of full light curves, especially flaring M dwarfs and pulsating stars. This is the reason we limit our parent sample to \textit{BP$-$RP}$=0.7-1.6$\,mag to avoid the pulsating stars on the bluer side and the flaring stars on the redder side \citep{Eyer2019}. Our color selection may still include some pulsating $\gamma$ Doradus variables, but most of $\gamma$ Doradus stars have fractional variability $<1\%$ \citep{Balona2011}, and are unlikely to dominate our EB sample. Our visual inspection of light curves also rules out $\gamma$ Doradus because $\gamma$ Doradus light curves tend to have long-term modulation due to its multiple pulsating modes \citep{Balona2011}.
    
    For our double eclipsing binary pairs (2+2 systems), we use the light curves from Wide-field Infrared Survey Explorer ({\it WISE}), the Zwicky Transient Facility (ZTF), and All-Sky Automated Survey for Supernovae (ASAS-SN) to confirm their eclipsing binary nature and measure their periods. {\it WISE} is an all-sky infrared survey \citep{Wright2010, Mainzer2014}, and we  use its $W1$-band ($3.4$\,$\mu$m) light curves for its better sensitivity and point spread function ($\sim6$\,arcsec) compared to its other bands. Light curves from {\it WISE} have been used to identify tens of thousands of eclipsing binaries \citep{Chen2018,Petrosky2021}. The Zwicky Transient Facility (ZTF) is an optical ground-based survey of astrophysical variability based at the Palomar Observatory \citep{Bellm2019}. ZTF offers an improved point spread function ($\sim2$\,arcsec) and depth (down to $\sim21$\,mag in $g$-band) over {\it WISE} at the expense of the sky coverage. We use ZTF $g$-band light curves when they are available, and use {\it WISE} if the source is not covered by ZTF. Finally, we also utilize ground-based optical ASAS-SN \citep{Shappee2014,Kochanek2017} data in cases where WISE and ZTF lack necessary spatial or temporal coverage. 
     
     We measure periods from the light curves using the Lomb-Scargle periodogram implemented in Astropy \citep{Astropy2013,Astropy2018}. We search for periodic signals from 0.1 to 10 day, with a frequency spacing of $10^{-4}$\,day$^{-1}$. For short-period eclipsing binaries, the peak in the periodograms typically represents half of the actual orbital period of binaries because their primary and secondary eclipses are difficult to distinguish. We thus phase-fold the light curves using the orbital period (i.e. two times the period peaking in the periodogram) and report the orbital periods if the phase-folded light curves look consistent with EB light curves \citep{Soszynski2016}. 
     
    \begin{figure*}[t]
    \centering
        \includegraphics[height=0.33\linewidth,valign=c]{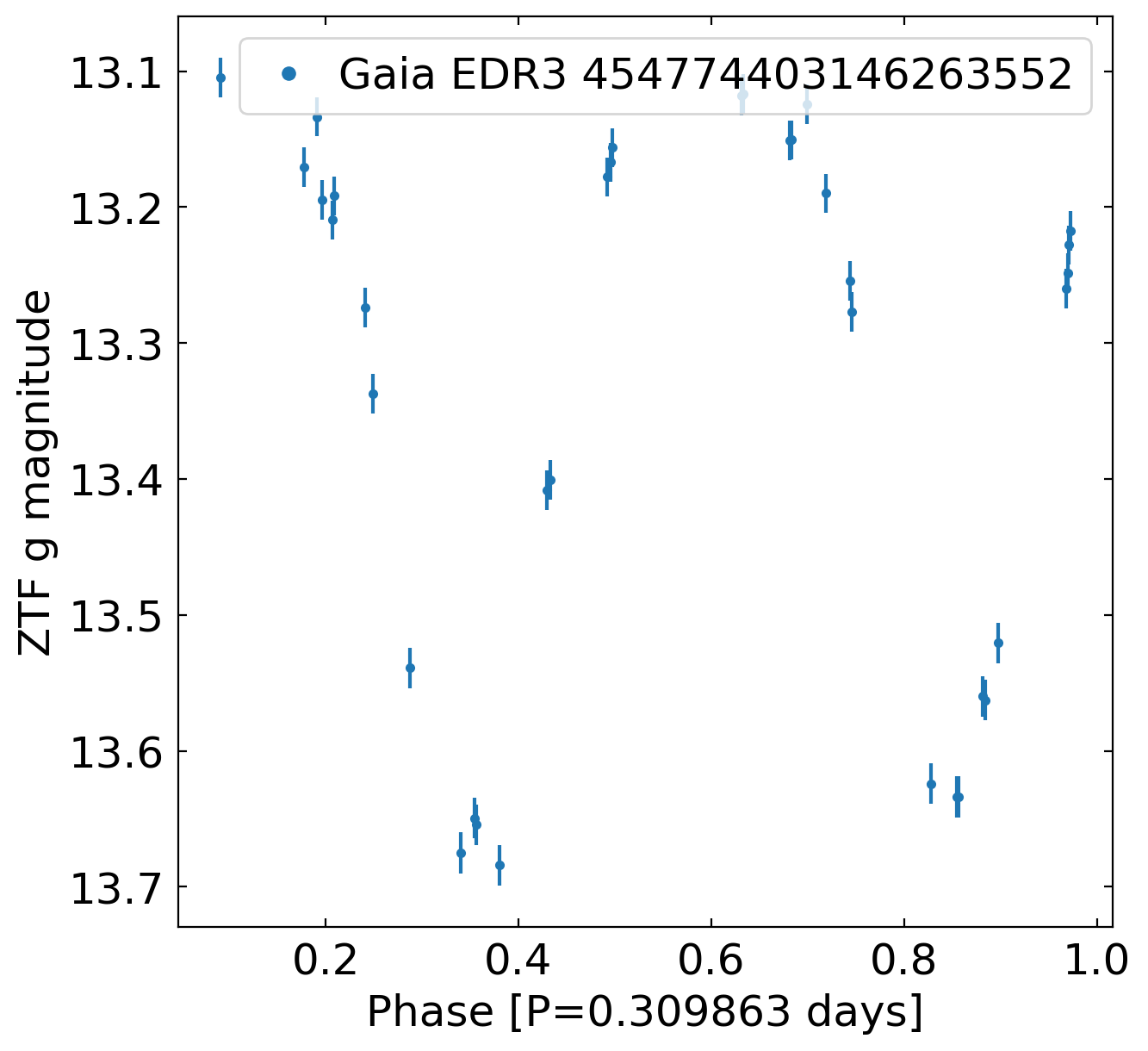}
        \includegraphics[height=0.25\linewidth,valign=c]{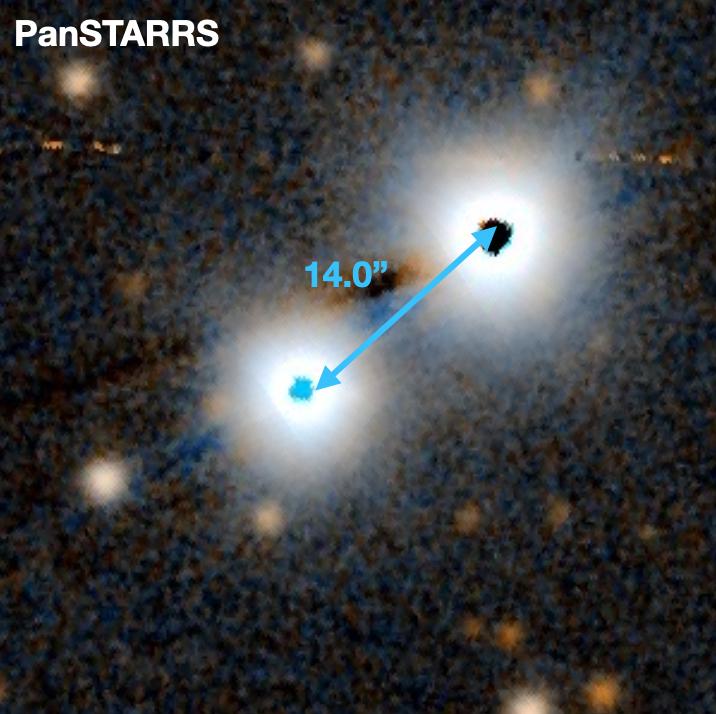}
        \includegraphics[height=0.33\linewidth,valign=c]{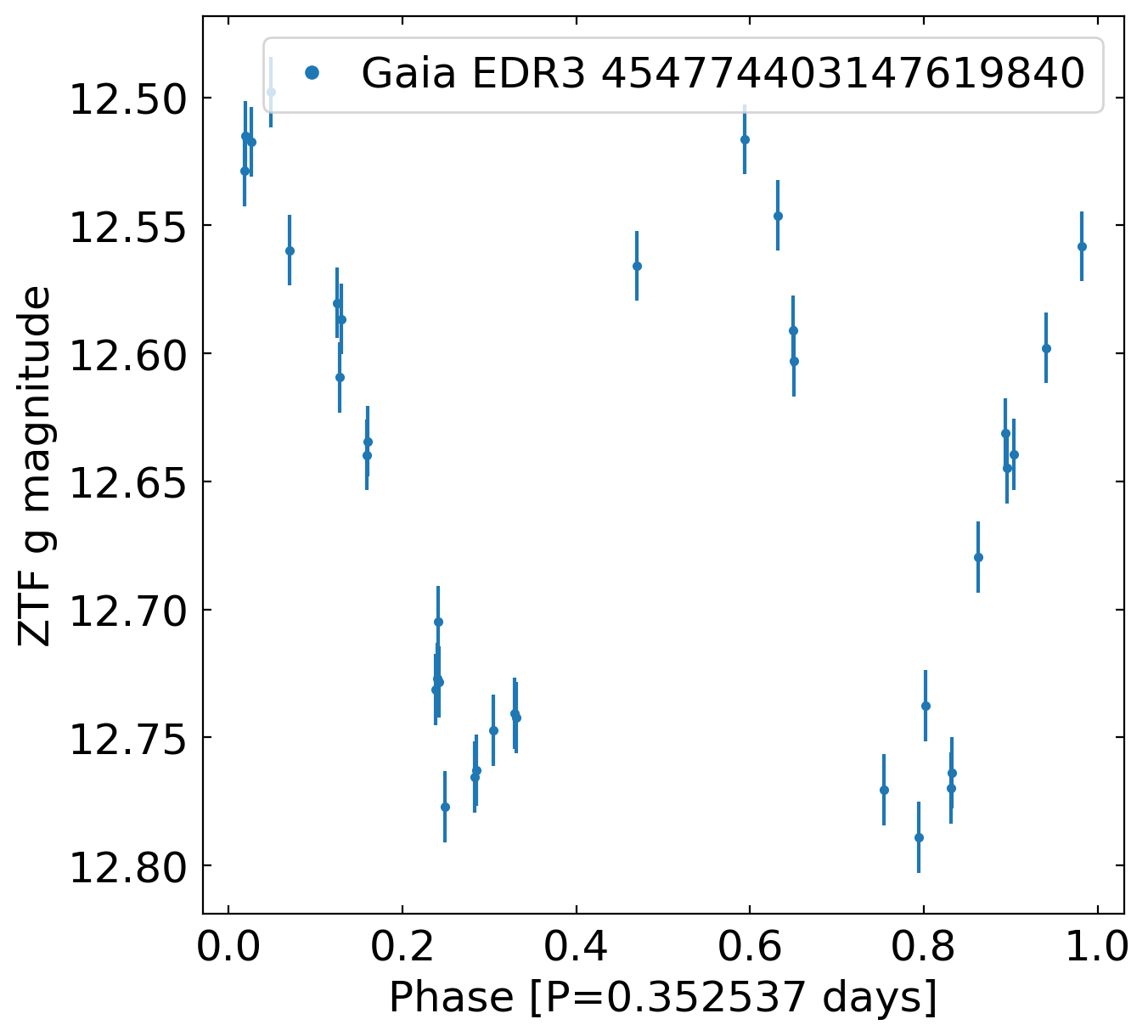}
        \includegraphics[height=0.33\linewidth,valign=c]{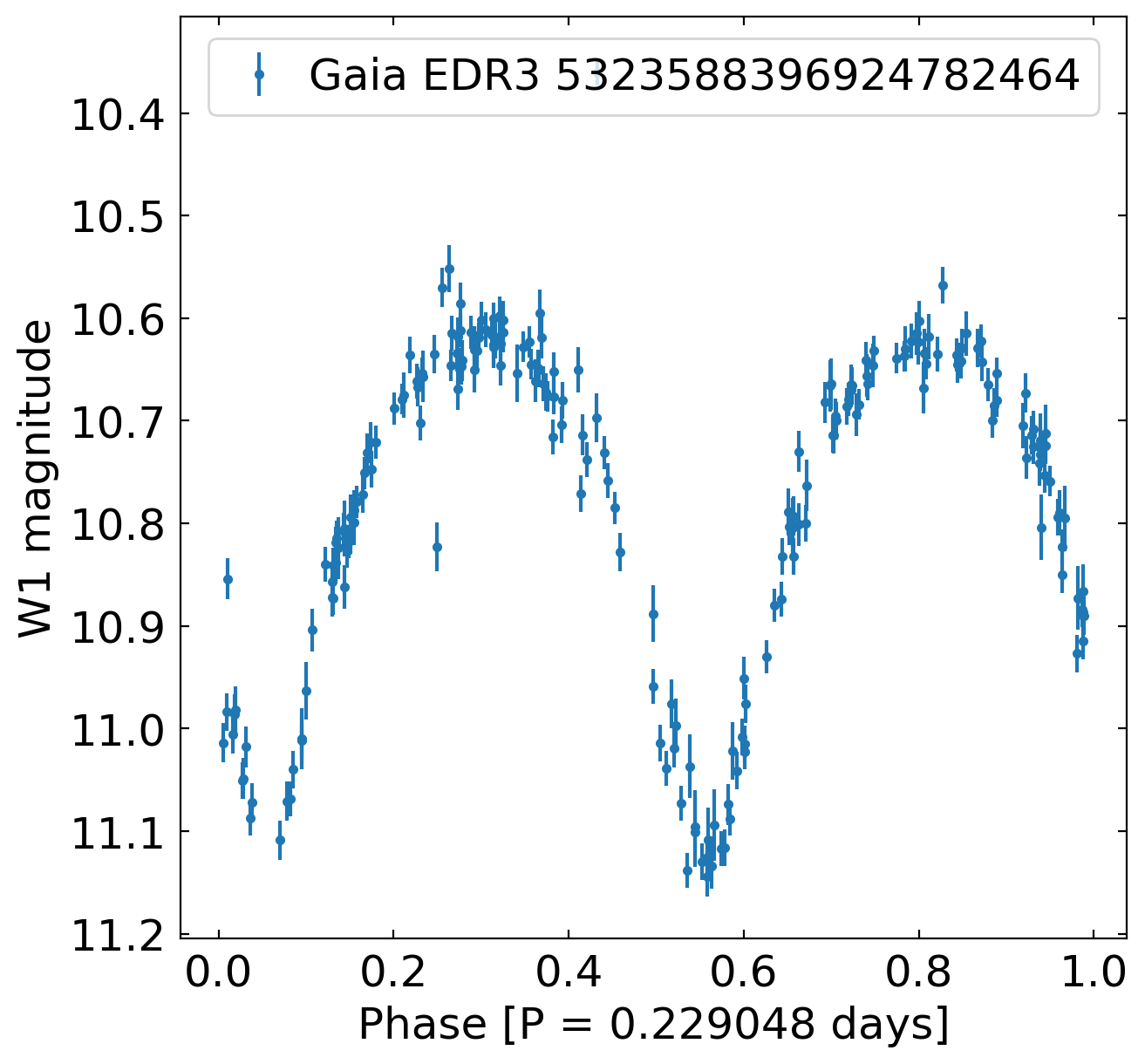}
        \includegraphics[height=0.25\linewidth,valign=c]{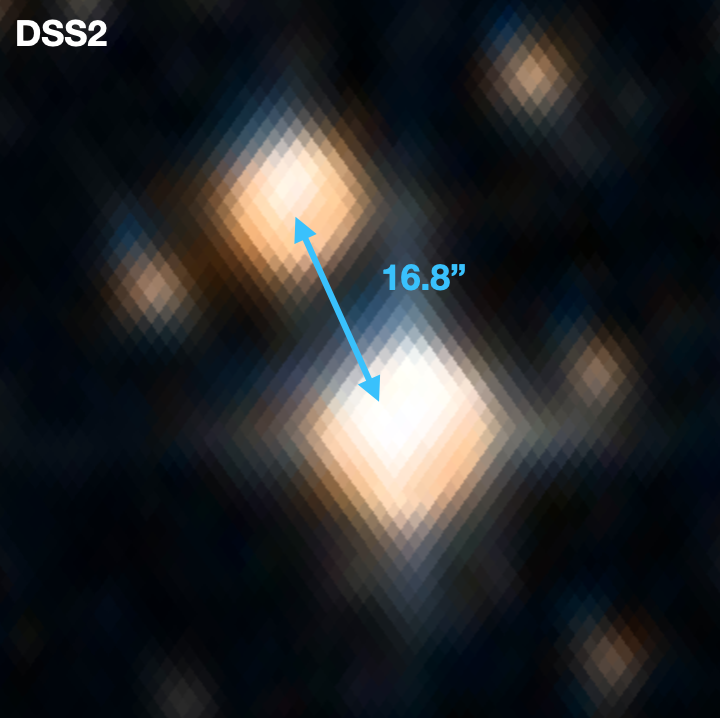}
        \includegraphics[height=0.33\linewidth,valign=c]{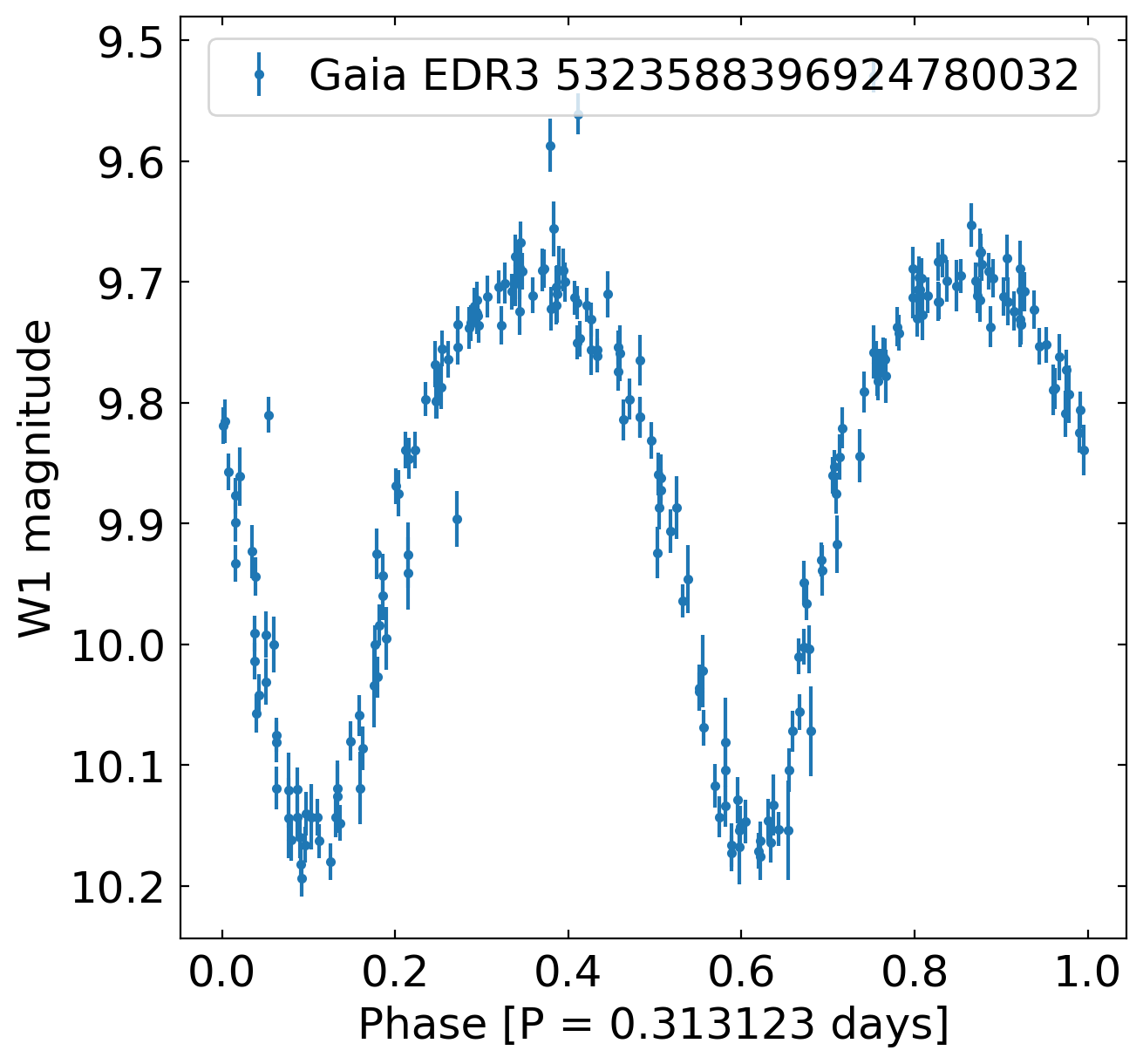}
        \includegraphics[height=0.33\linewidth,valign=c]{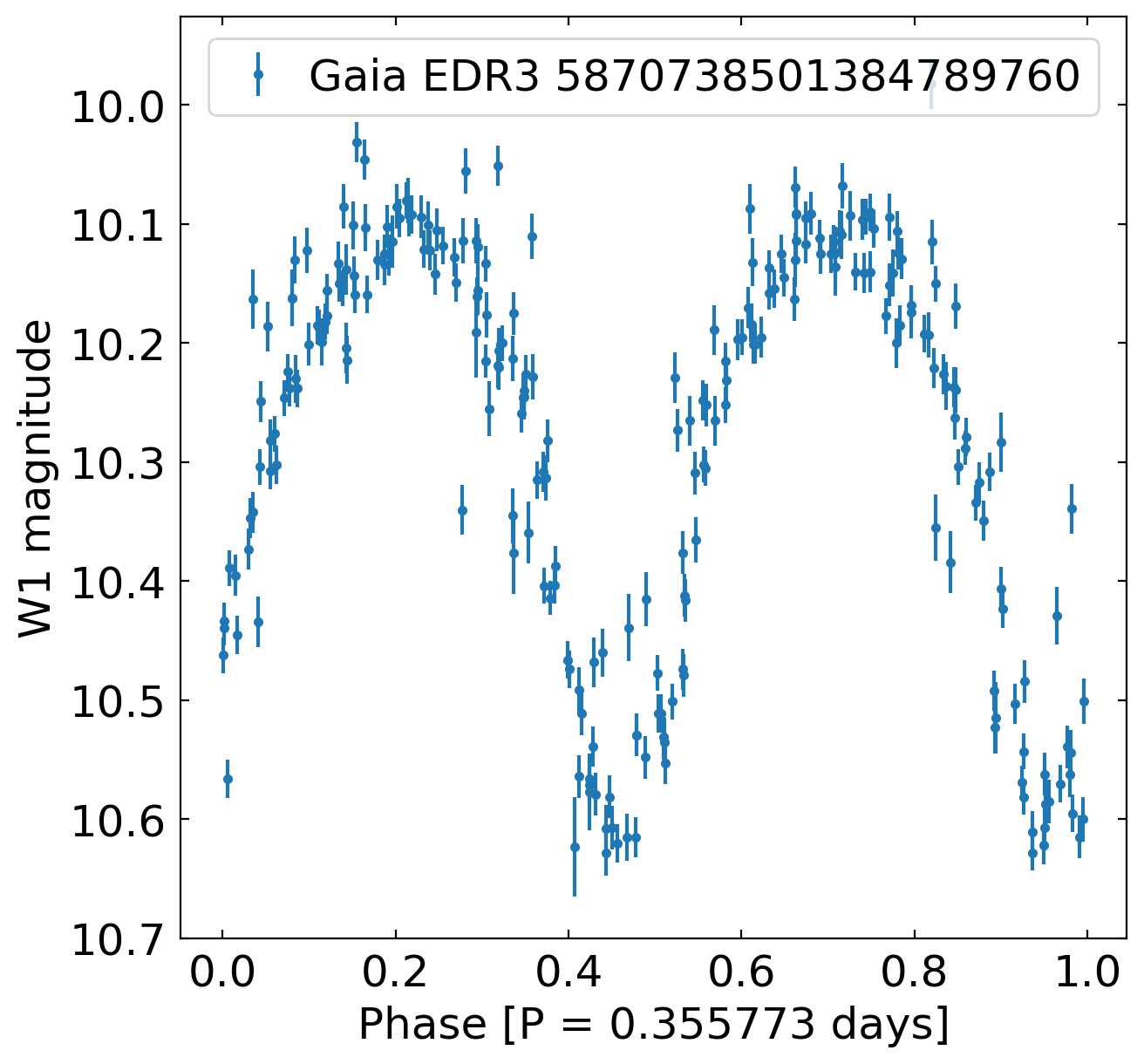}
        \includegraphics[height=0.25\linewidth,valign=c]{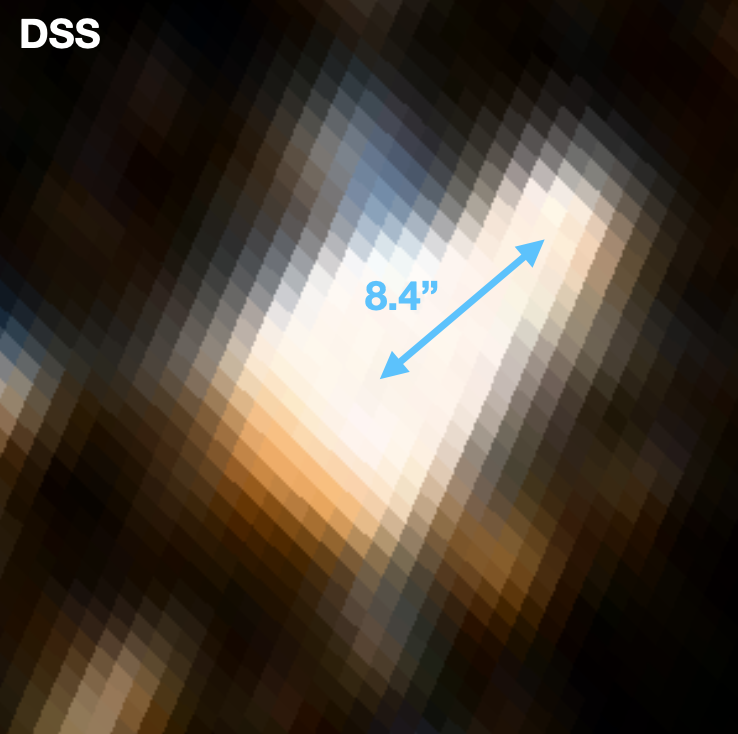}\hspace{3em}%
        \includegraphics[height=0.3\linewidth,valign=c]{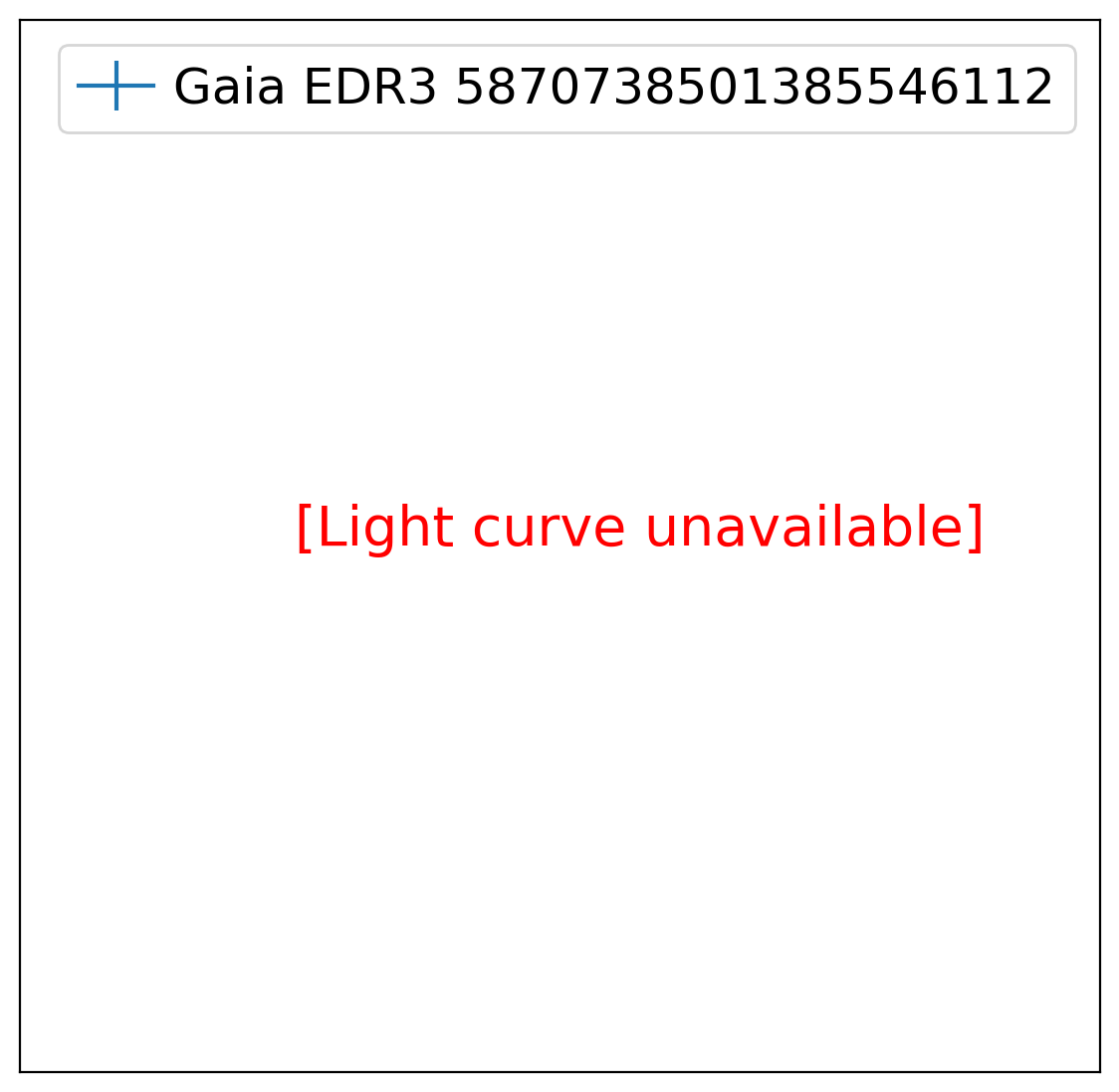}
    \caption{2+2 quadruples investigated using {\it WISE} $W1$-band or ZTF $g$-band photometric data, phase-folded by their orbital periods. Center: optical images from the DSS2 \& PanSTARRS surveys with the angular separation between the EBs labeled. Left and Right: the phase-folded light curves for each EB with periods listed at the bottom. Light curves are organized according to the stars' locations in the images. {\it Gaia} EDR3 5870738501385546112 is too blended with the brighter companion and has no available lightcurve as described in the text. \label{fig:lightcurves1}}
    \end{figure*}
    
    \begin{figure*}[t]
    \ContinuedFloat
    \centering
        \includegraphics[height=0.33\linewidth,valign=c]{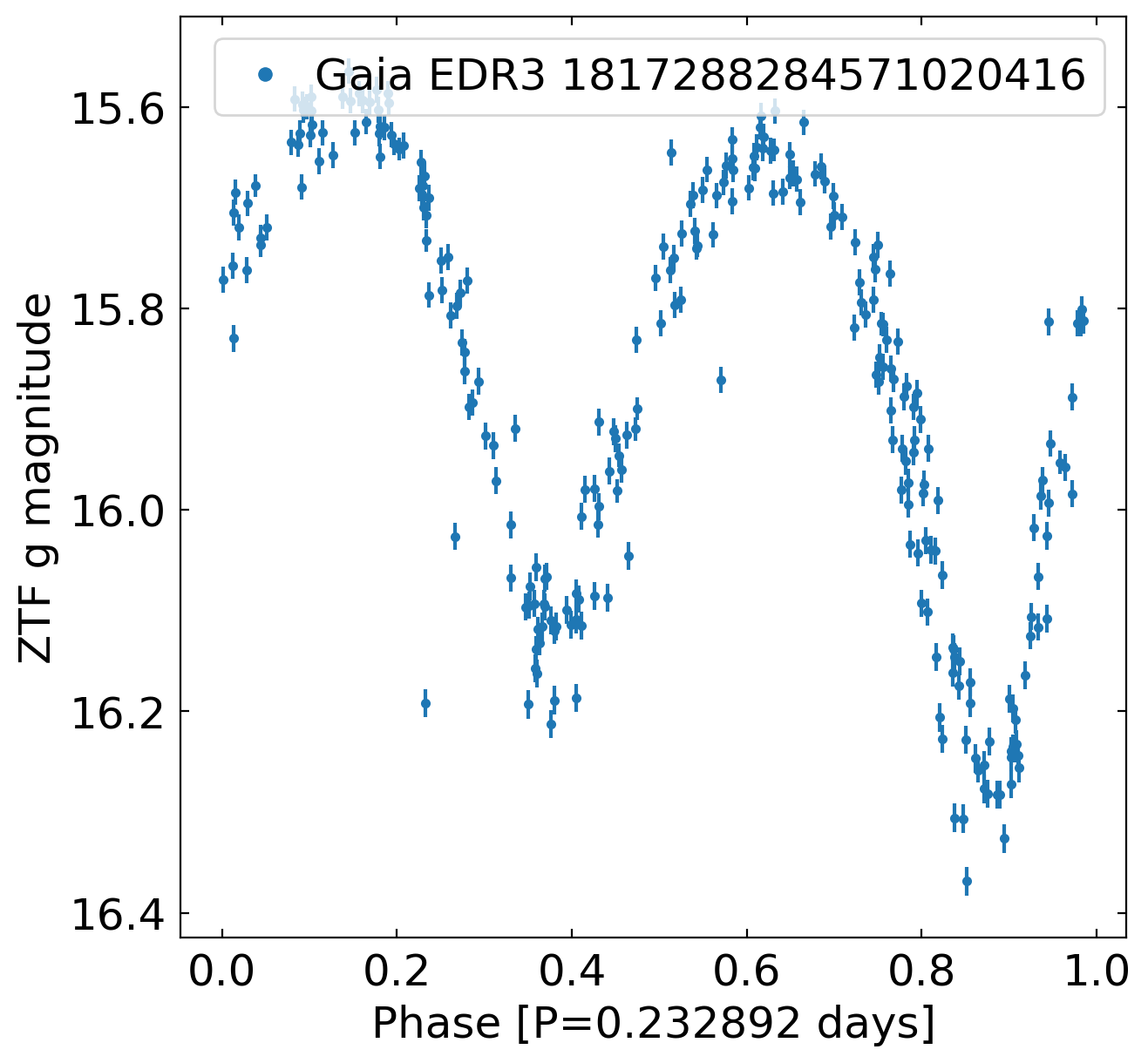}
        \includegraphics[height=0.25\linewidth,valign=c]{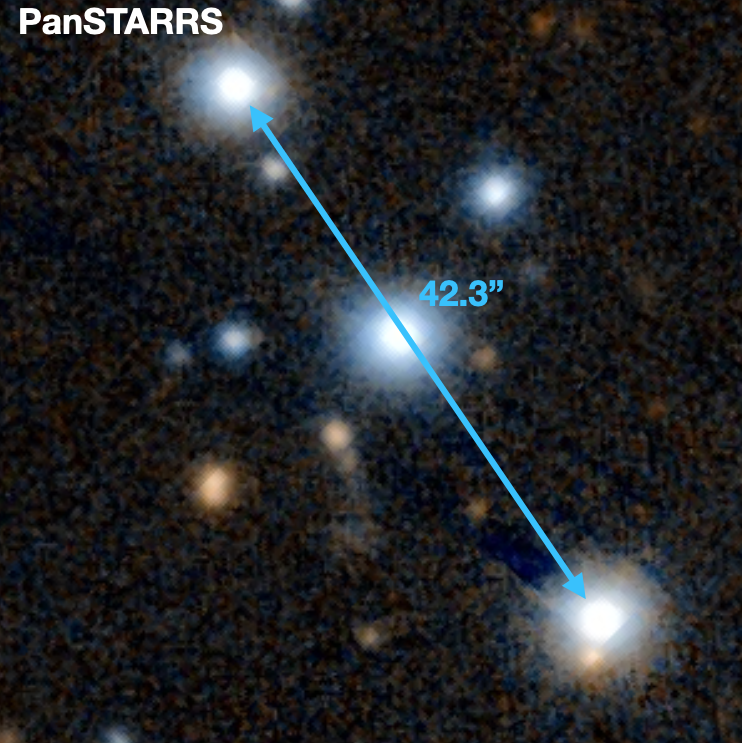}
        \includegraphics[height=0.33\linewidth,valign=c]{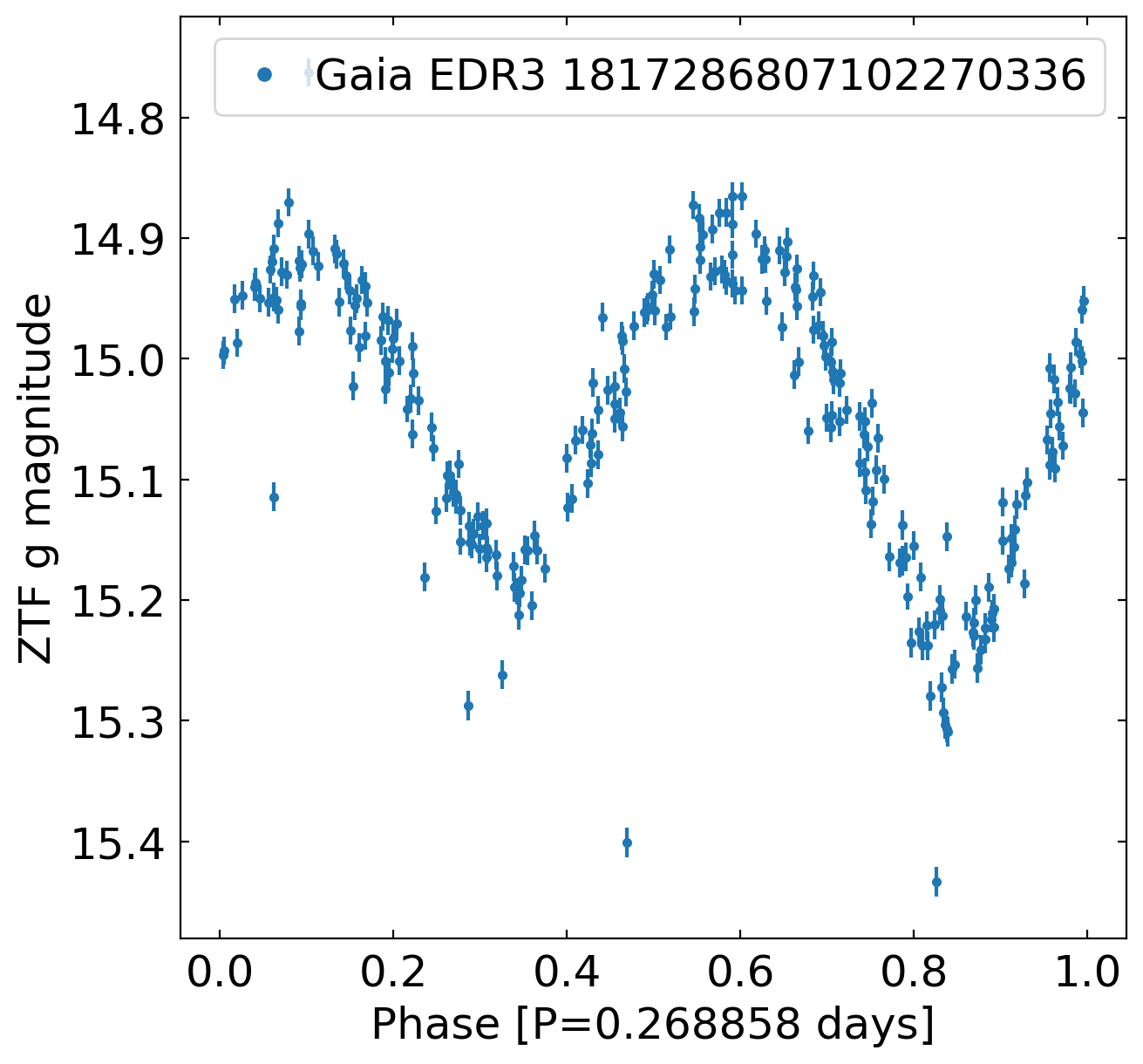}
        \includegraphics[height=0.33\linewidth,valign=c]{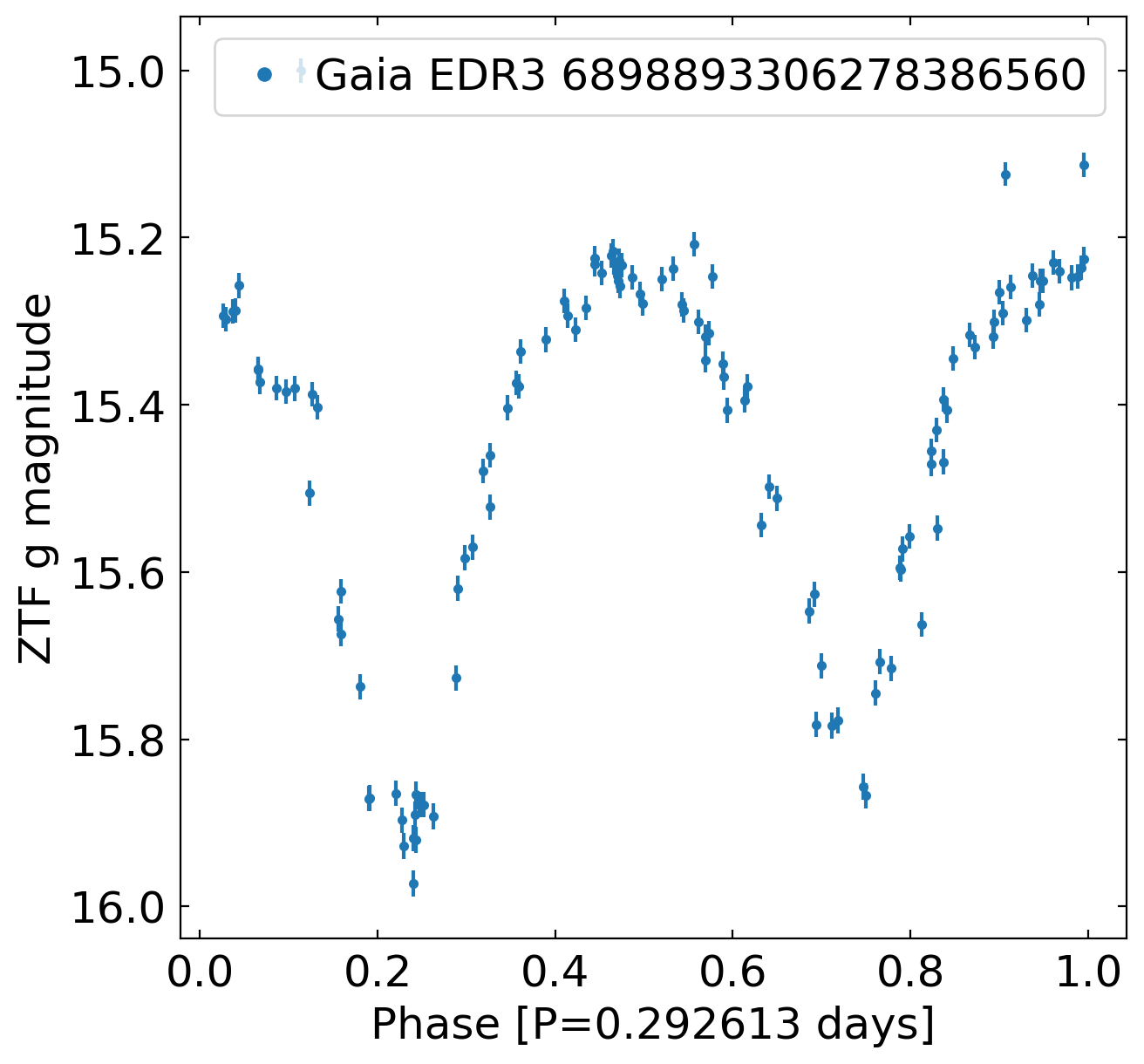}
        \includegraphics[height=0.25\linewidth,valign=c]{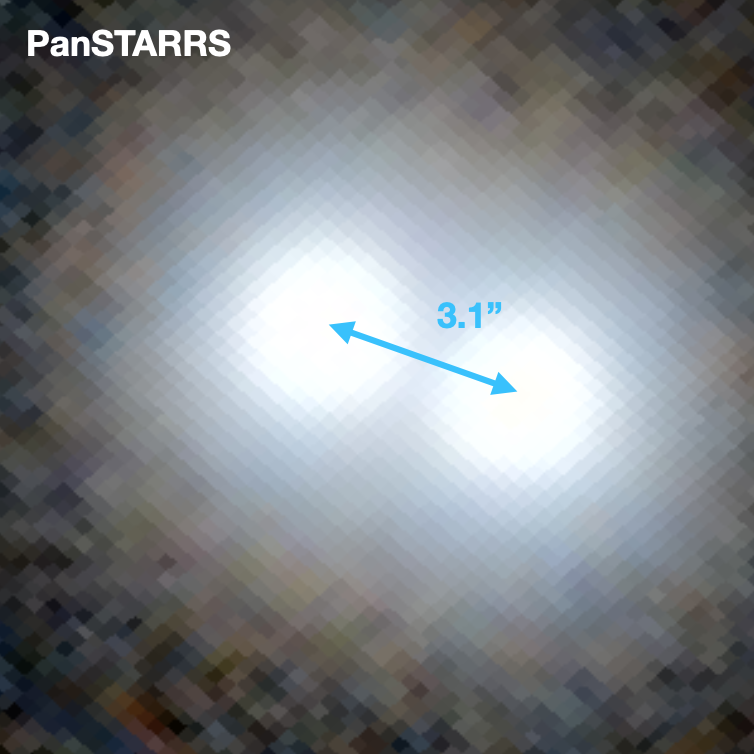}
        \includegraphics[height=0.33\linewidth,valign=c]{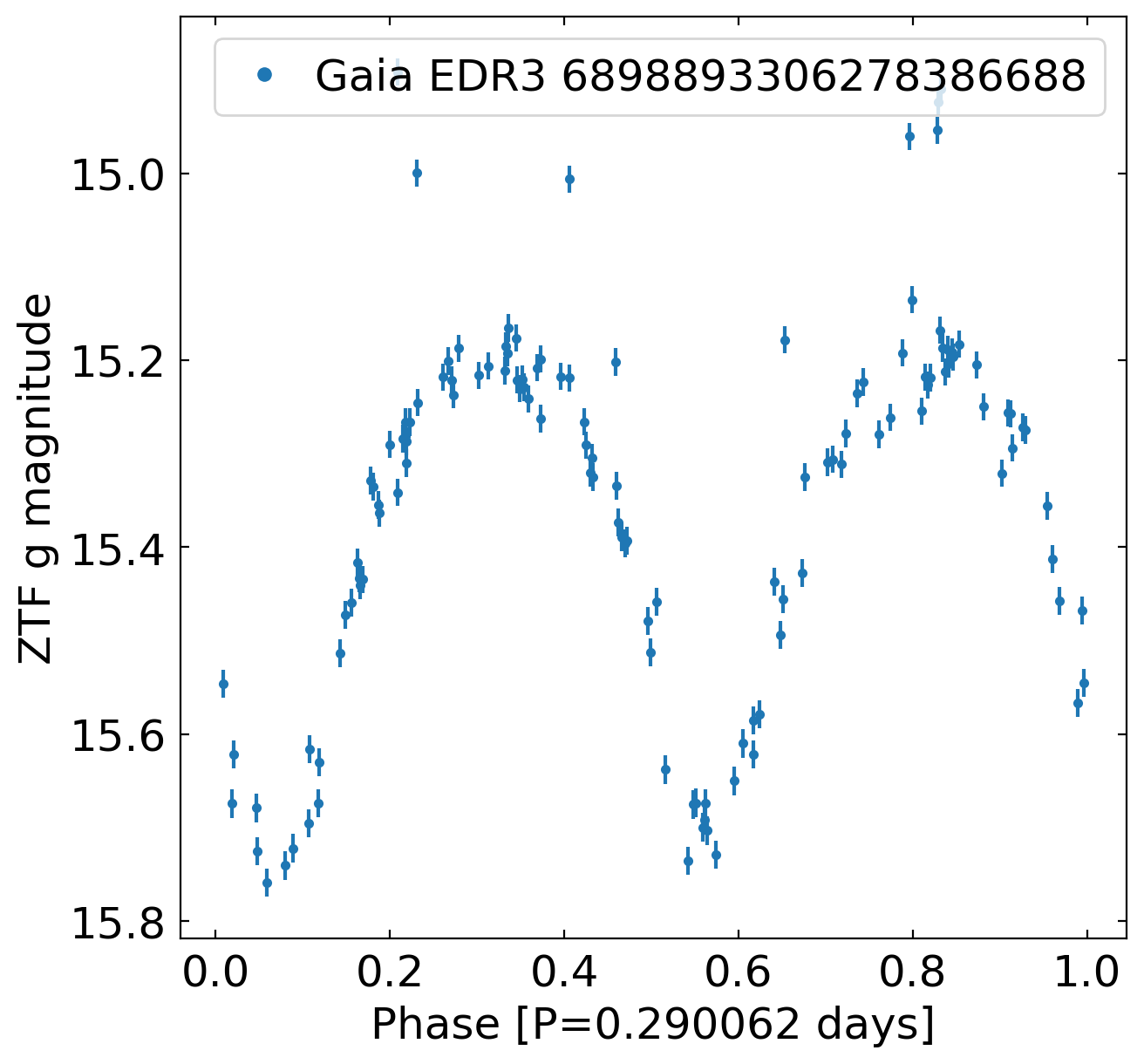}
        \includegraphics[height=0.33\linewidth,valign=c]{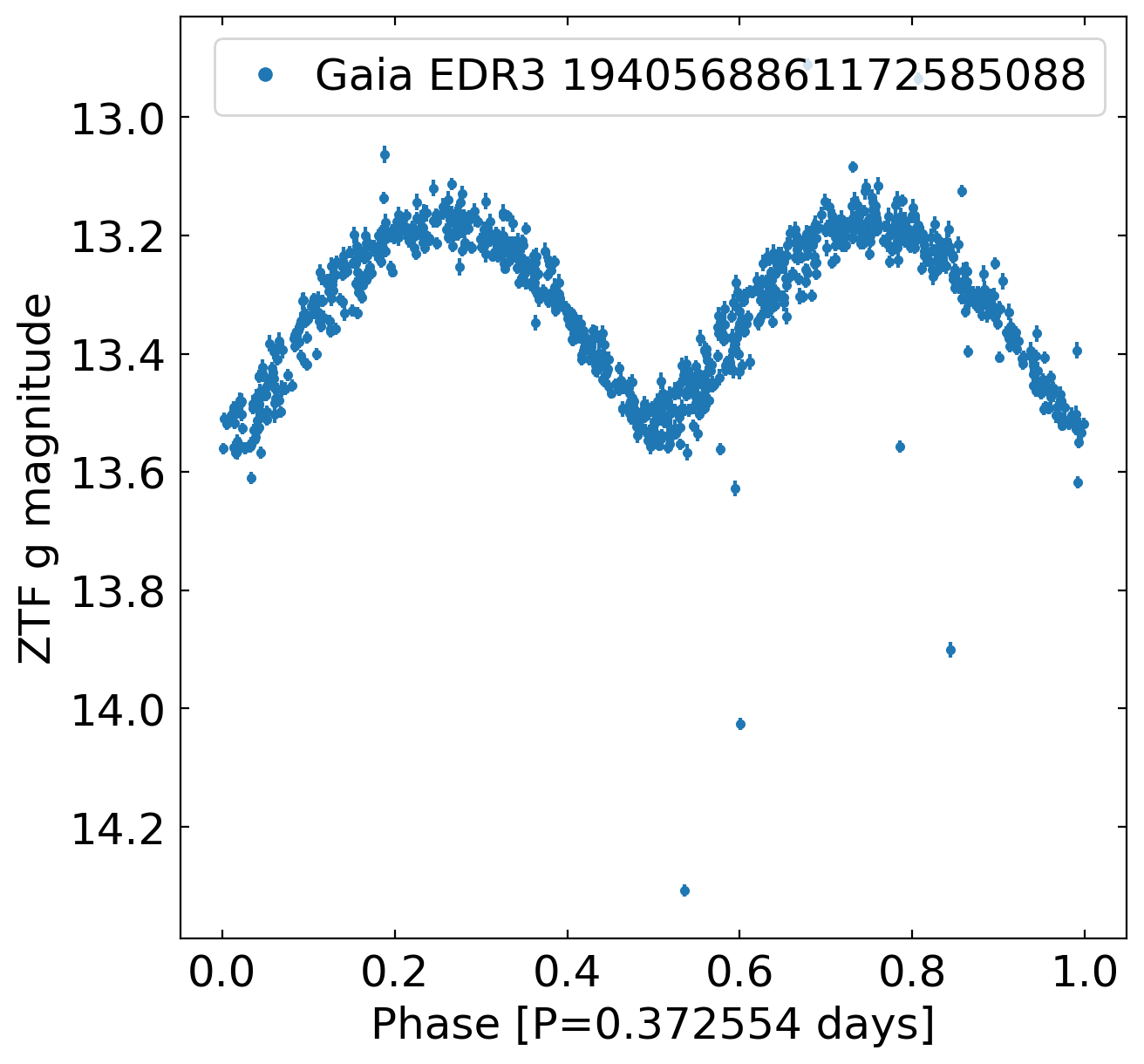}
        \includegraphics[height=0.25\linewidth,valign=c]{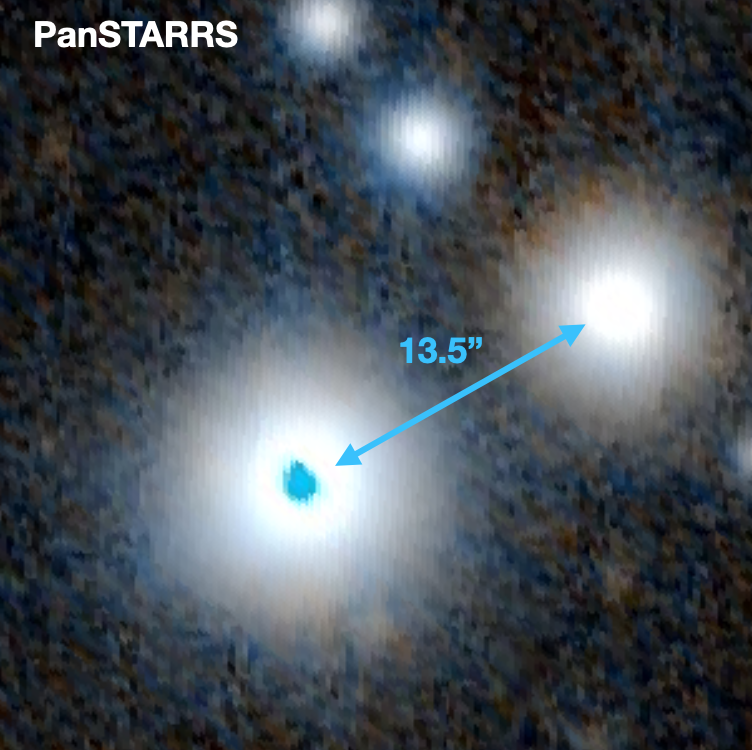}
        \includegraphics[height=0.33\linewidth,valign=c]{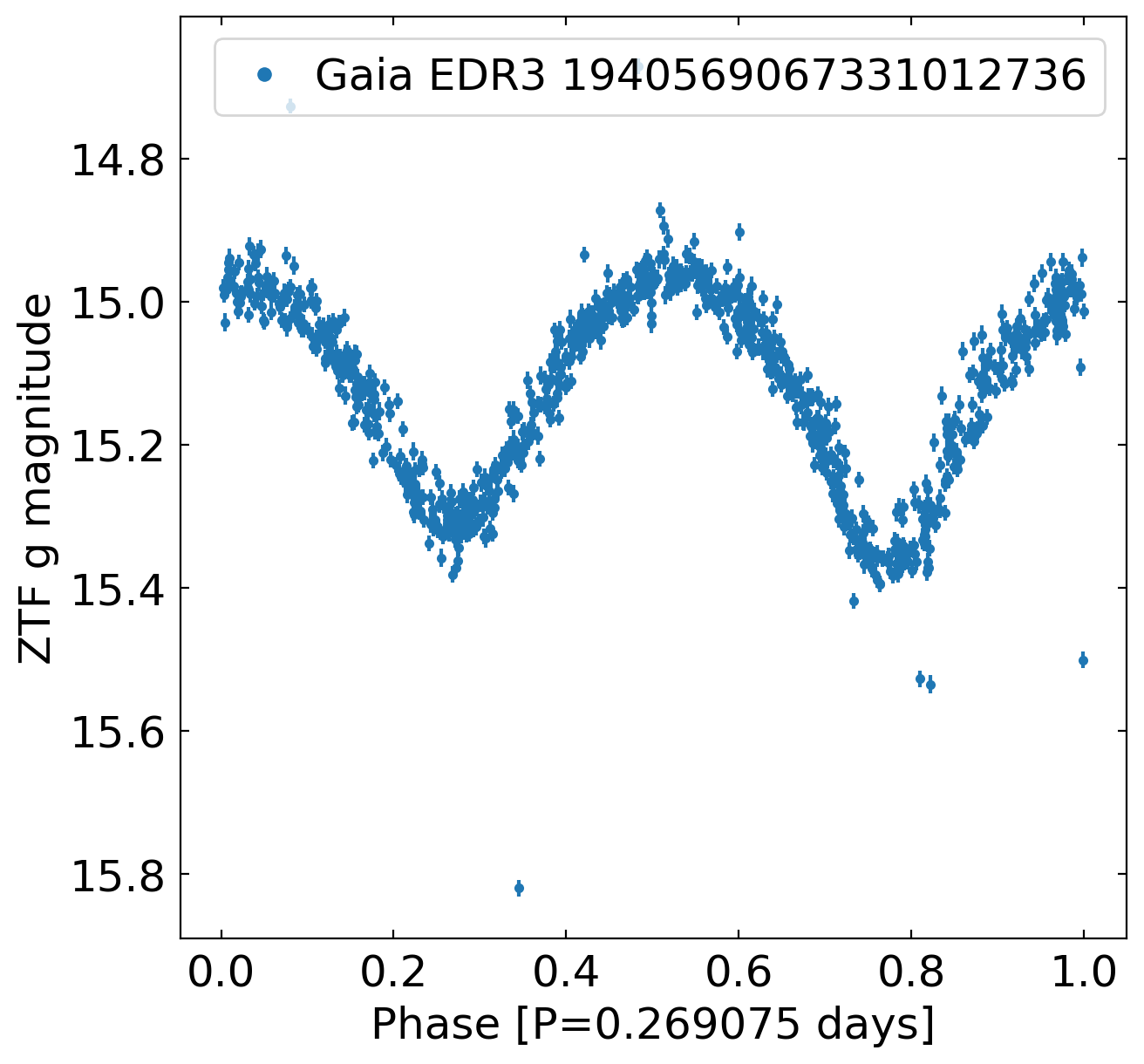}
    \caption{(Continued)\label{fig:lightcurves2}}
    \end{figure*}
    
\section{Results}
\label{sec:results}
    
\subsection{Light curve analysis of 2+2 quadruple systems}

    We conduct a full comoving search for all {\it Gaia} sources in the parent sample. Among $N_{MS}=15,684,999$ stars in the parent sample, we have $N_{EB}=26,734$ EBs selected by their $>$5\%\ variability. The comoving search shows that $N_{2+n}=1,282$ EBs have comoving companions, of which there are $8$ pairs of double-EB quadruples (2+2 systems), and thus $N_{2+2}=16$ EBs in 2+2 systems. We use the subscript $2+n$ to refer to the EBs that have comoving companions, regardless of whether the companion is eclipsing or not. Fig.~\ref{fig:sepvel2} shows the locations of these $2+n$ (open gray circles) and  $2+2$ (red diamonds) pairs in the relative velocity-separation space. Information regarding these 2+2 systems is tabulated in Table~\ref{tab:catalog5}.
    
    Seven out of the eight 2+2 systems are newly discovered. Our search recovers one known 2+2 system, BV Draconis and BW Draconis (pair {\it Gaia} EDR3 1616589410627042560 \& 1616589410627042688), supporting our methodology. The periods of these components in Table \ref{tab:catalog5} are obtained from a previous study \citep{Yamasaki1979}, and we do not re-measure the periods of these sources using modern data because they are too bright and suffer from saturation in the surveys. The two component binaries, initially discovered independently due to their variability, were then hypothesized to form a 2+2 quadruple based on similar photometric parallaxes, proper motions and systemic radial velocities \citep{Batten1986}. We now confirm the 2+2 quadruple nature of this system using high-quality {\it Gaia} parallaxes. 
    
\begin{figure}[t!]
        \centering
        \includegraphics[width = 0.95\linewidth]{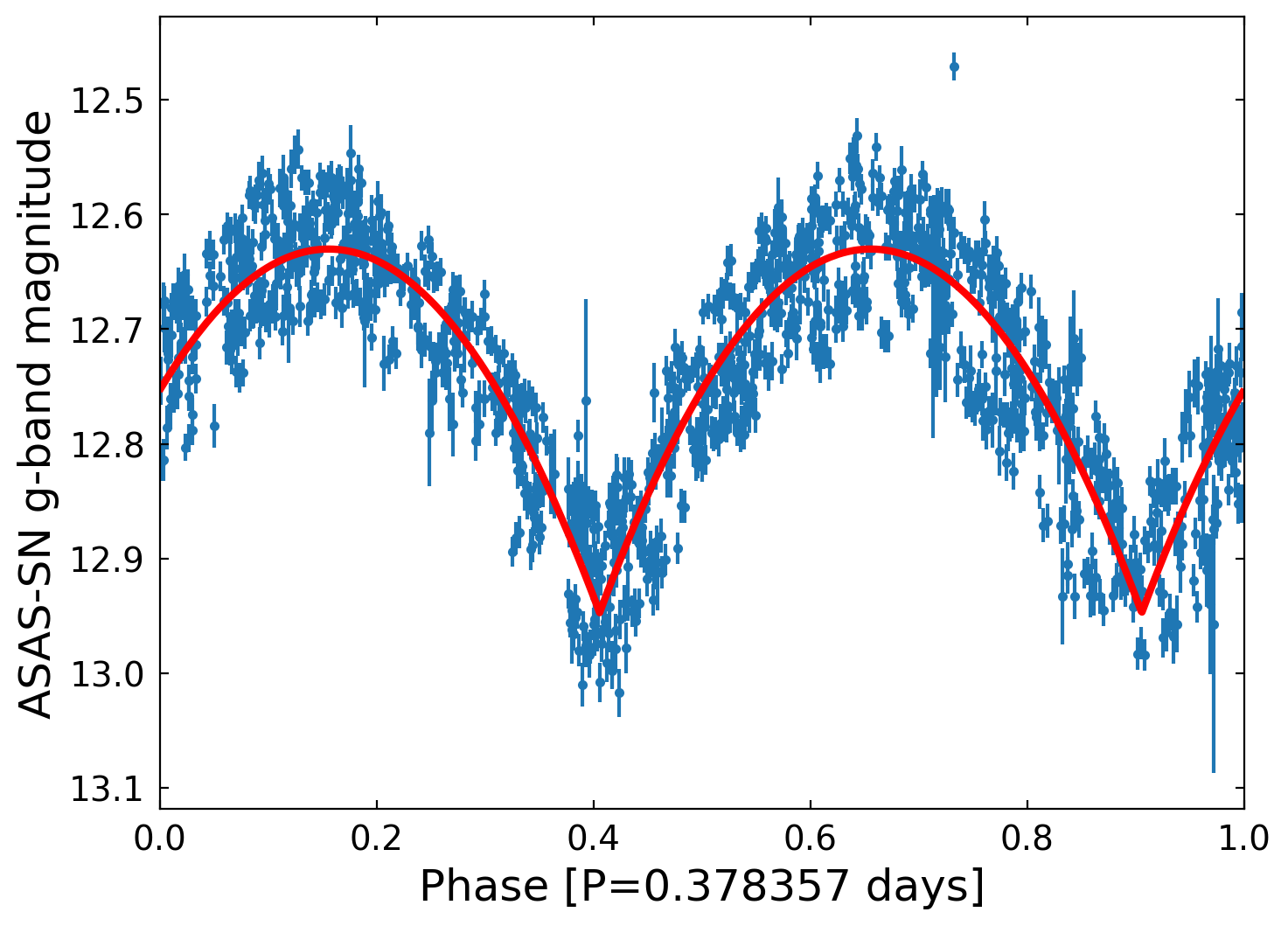}\\
        \includegraphics[width = 0.95\linewidth]{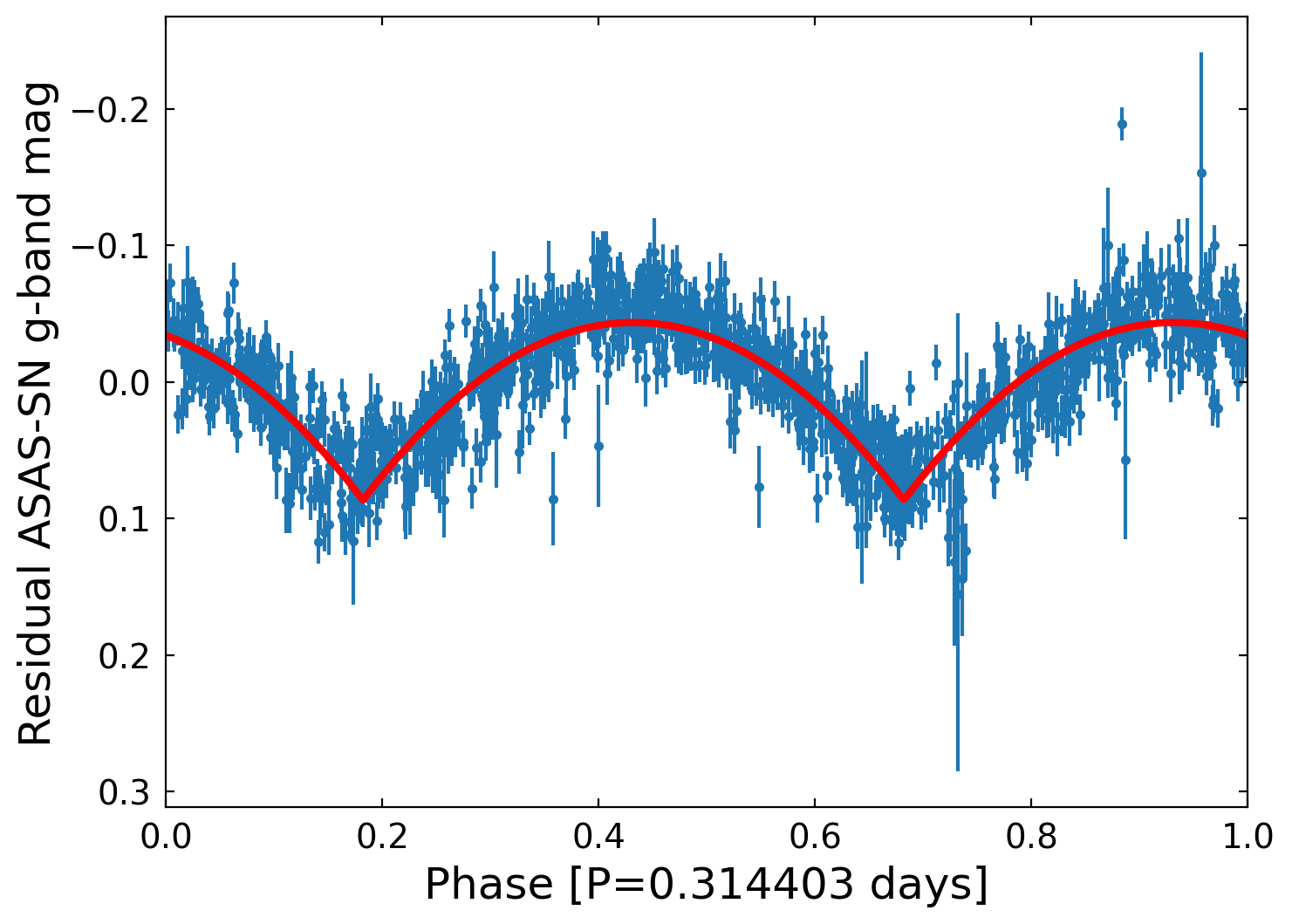}\\
        \includegraphics[width = 0.95\linewidth]{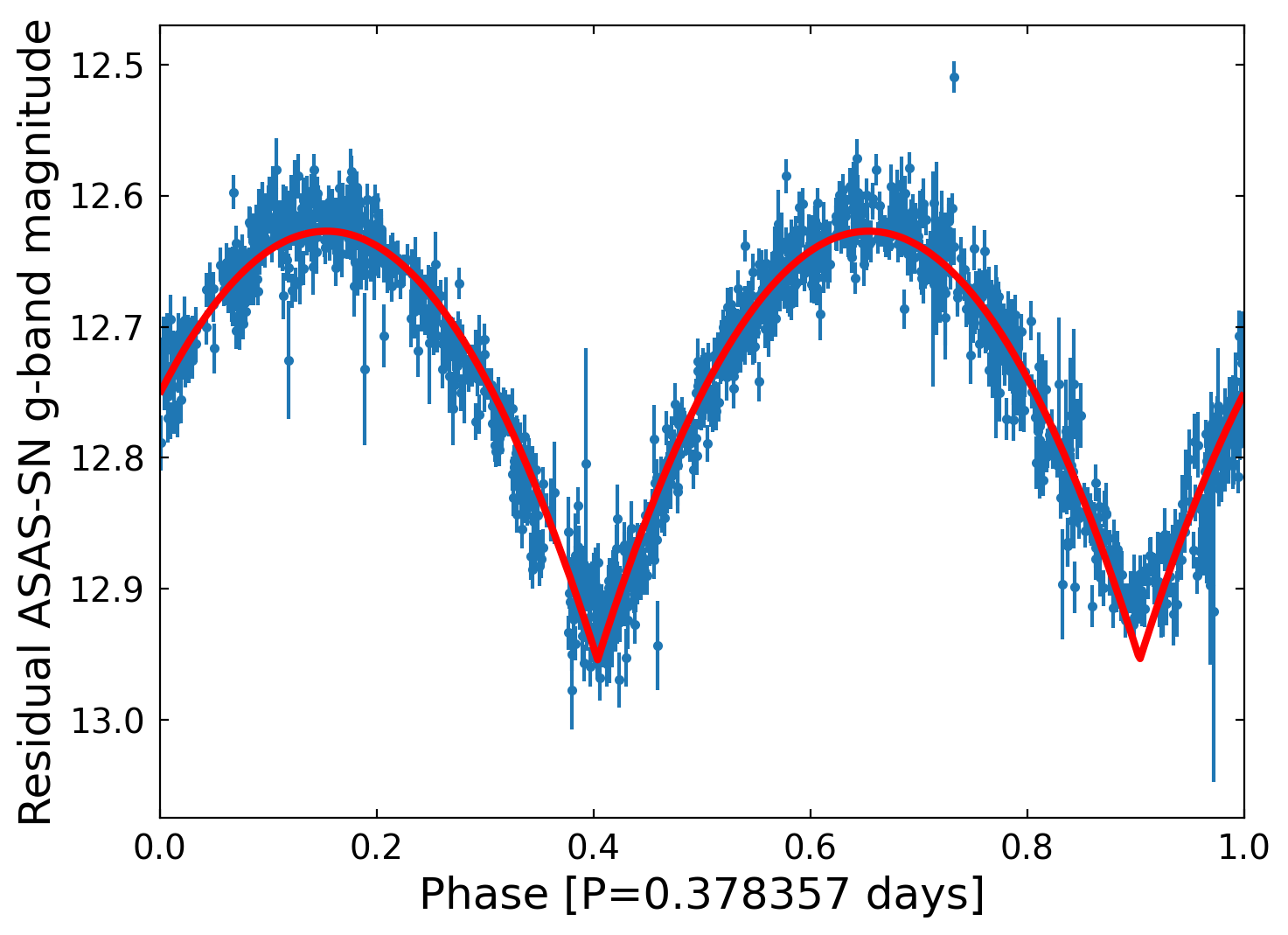}\\
        \caption{The deblending procedure for the pair {\it Gaia} EDR3 6724245225761139456 and 6724245230088060416. Top: The total blended ASAS-SN $g$-band light curve phase-folded to the strongest periodicity. The fitted piece-wise parabolic model is shown in red. Middle: The residual light curve after the best model is subtracted from the total light curve. It is then fitted with a new parabolic model shown in red. This secondary variability is most likely from the fainter component. Bottom: The phase-folded light curve of the brighter component after the variability of the fainter component is subtracted from the total light curve using the model from the middle panel.}
        \label{fig:deblend}
    \end{figure}

    We perform periodogram analysis for all fourteen EBs in the seven new 2+2 systems. The components in all seven 2+2 pairs are well-separated by $>3$ arcsec, and we successfully measure the orbital periods for 13 out of 14 EBs from their light curves in Fig.~\ref{fig:lightcurves1} and Fig.~\ref{fig:deblend}. Fig.~\ref{fig:lightcurves1} shows light curves of individual EB components folded by orbital periods measured from the periodograms, with optical images from Pan-STARRS or DSS if Pan-STARRS is not available. We place light curves so that the left panel corresponds to the left star in the optical image (middle panel), and the right panel corresponds to the right one. We catalog the information about these 2+2 systems in Table~\ref{tab:catalog5}, including {\it Gaia} EDR3 source IDs, component coordinates, and orbital periods for each EB, as well as their projected separations. The fact that the phase-folded light curves are consistent with EB classification (e.g., \citealt{Soszynski2016}) suggests that our selection using {\it Gaia} 5\% variability is effective in selecting EBs with low contamination. 
    
    The pair {\it Gaia} EDR3 6724245225761139456 and 6724245230088060416 is not covered by ZTF, and the {\it WISE} light curves are not reliable due to the pair's small angular separation of 8.5\,arcsec, which is only marginally larger than the PSF of {\it WISE}. Instead, we obtain its light curves from ASAS-SN. Although ASAS-SN does not spatially resolve two individual components due to its large photometric aperture ($\sim16$\,arcsec, \citealt{Kochanek2017}), the two components in this pair have comparable $G$-band magnitudes (12.71 and 13.35\,mag, respectively), and thus both components contribute their variability to the blended ASAS-SN light curve. 
    
    To deblend the light curve, we first run the periodogram on the light curve, resulting in a peak period of 0.378/2\,day (Fig.~\ref{fig:deblend}, top panel). Then we fit a piece-wise parabolic model at the period of 0.378\,day (red solid line) and subtract the best fit from the ASAS-SN light curve to obtain the residual light curve. We measure the periodogram for the residual light curve, resulting in another peak period of 0.314/2\,day (middle panel). We then fit another piece-wise parabolic model to the middle panel and subtract the best fit from the total light curve (bottom panel). Even though the dominant period is well apparent in both the top and the bottom panels, compared to the top panel, the bottom panel where the sub-dominant variability has been subtracted has significantly reduced residuals. Given that the cadence of ASAS-SN is $\sim1$\,day, the period difference of 0.064\,day is not due to aliasing, suggesting that these two periods are physical. Their amplitude radio is $\sim2$, similar to their G-band flux ratio of 1.8, suggesting that the larger-amplitude light curve with an orbital period 0.378\,day is coming from the brighter source. 
    
    We are not able to measure the period of one EB ({\it Gaia} EDR3 5870738501385546112) because it is not covered by ZTF, and its {\it WISE} light curve is dominated by its brighter companion (brighter by 2.92\,mag in $G$ band) 8 arcsec away, which is only slightly larger than the PSF of {\it WISE}. Therefore, we only present the {\it WISE} light curve for its bright companion in Fig.~\ref{fig:lightcurves1}. We attempt the same deblending procedure for the joint ASAS-SN light curve from the pair EDR3 5870738501385546112 and 5870738501384789760, but we still do not detect a promising eclipsing signature in the residual, most likely because the second component is much fainter than the dominant one. The brighter component {\it Gaia} EDR3 5870738501384789760 is a known eclipsing binary (V* OZ Cen), presented with a period measurement of 0.355773710 day from \citet{Alfonso-Garzon2012}, in agreement with our measurement. 
    
    Although the angular separation of the pair {\it Gaia} EDR3 6898893306278386688 and 6898893306278386560 is only $3.1$\,arcsec, the ZTF light curves of the two components are spatially well-resolved. The measured orbital periods of the components look similar (0.290 and 0.293 day, respectively), but the difference is significant because the error of period determination is much smaller than the period difference of 0.003 day, for example $10^{-6}$\,day for {\it WISE}'s baseline and better than $10^{-5}$\,day for ZTF's baseline \citep{Petrosky2021}. 

    The pair {\it Gaia} EDR3 1817286807102270336, 1817288284571020416 is of particular interest as it is a possible resolved 2+2+1 system, with the two stars investigated here comoving with a third source, {\it Gaia} EDR3 1817286807098038656, which is about 2.5" from {\it Gaia} EDR3 1817286807102270336. This third source is red, with \textit{BP$-$RP}$=1.67$\,mag, and thus it is outside the color cut used in our sample, but it does meet the comoving search criteria outlined in Sec. \ref{sec:comoving}. The third source has a somewhat different parallax of $1.908\pm0.174$\,mas, compared to $1.6795\pm0.0199$ and $1.6360\pm0.0263$\,mas of the other two sources, but the difference is only about 1.5 $\sigma$ due to the large parallax uncertainty of the third source.
    
    The inner periods of our 2+2 quadruple candidates are very short and range between 0.2 and 0.4 days. Variability-selected EB samples are dominated by short-period (contact and semi-detached) binaries \citep{Paczynski2006}. This is due largely to selection biases, since at smaller separations both the chance of a suitable alignment with the line of sight is greater and the fraction of time that the binary spends in an eclipse is greater. Furthermore, contact and semi-detached binaries have larger root-mean-squared variability compared to detached binaries at the same periods. Indeed, the phase-folded light curves of our EBs in Fig.~\ref{fig:lightcurves1} are all consistent with them being contact or near-contact binaries.  
    
    %http://simbad.u-strasbg.fr/simbad/sim-id?Ident=V*+OZ+Cen&NbIdent=1
    
    \begin{table*}[t!]
        \centering
        \caption{Catalog of 2+2 quadruples.}
        \label{tab:catalog5}
        \begin{tabular}{lllllllll}
        \hline \hline
        Source ID & RA & Dec & G & Period & Separation & Separation \\
        & & & & (days) & (arcsec) & (AU) $\times$ 1000 \\
        \hline \hline
        454774403147619840 & 41.81221 & 56.76018 & 12.08 & 0.352537 & \multirow{2}{*}{14.0} & \multirow{2}{*}{4.72} \\
        454774403146263552 & 41.81775 & 56.75776 & 12.69 & 0.309863  \\
        \hline
        5323588396924782464\tablenotemark{c} & 134.70949 & -52.68316 & 11.48 & 0.313123 & \multirow{2}{*}{16.8} & \multirow{2}{*}{4.02} \\
        5323588396924780032 & 134.71295 & -52.67900 & 13.11 & 0.229048 &  & \\
        \hline
        5870738501385546112\tablenotemark{c} & 206.70398 & -59.26462 & 14.91 & --- & \multirow{2}{*}{8.4} & \multirow{2}{*}{3.21} \\
        5870738501384789760 & 206.70760 & -59.26605 & 11.99 & 0.355773 & &  \\
        \hline
        1616589410627042688\tablenotemark{a} & 227.95723 & 61.86188 & 8.68 & 0.292166\tablenotemark{a} & \multirow{2}{*}{16.0} & \multirow{2}{*}{0.99} \\
        1616589410627042560\tablenotemark{a} & 227.95828 & 61.85745 & 8.05 & 0.350066\tablenotemark{a} &  &  \\
        \hline
        6724245225761139456 & 270.08725 & -43.47736 & 13.35 & 0.314403\tablenotemark{b} & \multirow{2}{*}{8.5} & \multirow{2}{*}{5.48} \\
        6724245230088060416 & 270.08966 & -43.47578 & 12.71 & 0.378357\tablenotemark{b} & &  \\
        \hline
        1817286807102270336\tablenotemark{c} & 309.17715 & 20.12090 & 14.35 & 0.268858 & \multirow{2}{*}{42.3} & \multirow{2}{*}{25.54} \\
        1817288284571020416 & 309.18427 & 20.13058 & 15.02 & 0.232892 & & \\
        \hline
        6898893306278386688 & 319.18408 & -6.56223 & 14.89 & 0.290062 & \multirow{2}{*}{3.1} & \multirow{2}{*}{2.93} \\
        6898893306278386560 & 319.18488 & -6.56194 & 14.87 & 0.292613 &  &  \\
        \hline
        1940569067331012736 & 357.18219 & 48.82093 & 14.49 & 0.269075 & \multirow{2}{*}{13.5} & \multirow{2}{*}{9.10} \\
        1940568861172585088 & 357.18716 & 48.81908 & 12.94 & 0.372554 &  &  \\
        \hline \hline
        \end{tabular}
        \tablenotetext{a}{The known 2+2 systems, BV Draconis and BW Draconis, and we quote the orbital periods measured from \cite{Yamasaki1979}} \tablenotetext{b}{The deblended 2+2 system, where the orbital periods are assigned based on the brightness of stars.} \tablenotetext{c}{Comoving pairs that are not in \cite{El-Badry2021}. }
    \end{table*}
    
\subsection{Spectroscopic follow-up}

    We conducted spectroscopic follow-up of the system with {\it Gaia} IDs 1817286807102270336 and 1817288284571020416, whose light curves are shown in Figure \ref{fig:lightcurves2}, using the 3.5 m telescope at the Apache Point Observatory (APO). We observed the target on October 23, 2019 using Dual Imaging Spectrograph (DIS), with B1200 and R1200 gratings ($R\sim1500$) for the blue and the red camera and a slit width of 1.5 arcsec. We placed two eclipsing binaries (separated by 42\,arcsec) in the slit and observed the pair for 4.3 hours, split into 52 five-minute single exposures. We obtained arc observations every 30 minutes for reliable wavelength calibration. The raw data was flat-fielded and reduced using the standard IRAF pipeline. An unrelated star located midway between the two science targets (seen in Figure \ref{fig:lightcurves2}) provided an additional check on the wavelength stability. The measured velocity width of the absorption lines in the unrelated star taken to be due to instrumental resolution is $\sigma_{\rm ins}=40\pm 3$ \kms. 
       
    It was not possible to spectrophotometrically calibrate the datasets, both because of the weather conditions and because of a long-standing instrumental scattering issue with the DIS instrument. Therefore, instead of cross-correlating the complete spectra against stellar templates in order to determine radial velocities, rotational broadening, and radial velocity variations, we choose a limited wavelength range (6000-6200\AA) where there are several stellar absorption lines with relatively high equivalent widths. We then normalize the continuum by fitting a smooth polynomial to the relatively line-free spectral windows. We use high-resolution spectra of K and M dwarfs from \citet{Blanco2014}, downgraded to the spectral resolution of the instrument, to identify six highest equivalent width features in this spectral range on the air wavelength grid. For each binary in the 2+2 system, we then fit these features simultaneously with fixed relative amplitudes. 
    
\begin{figure}[h!]
    \centering
        \includegraphics[width=8.5 cm]{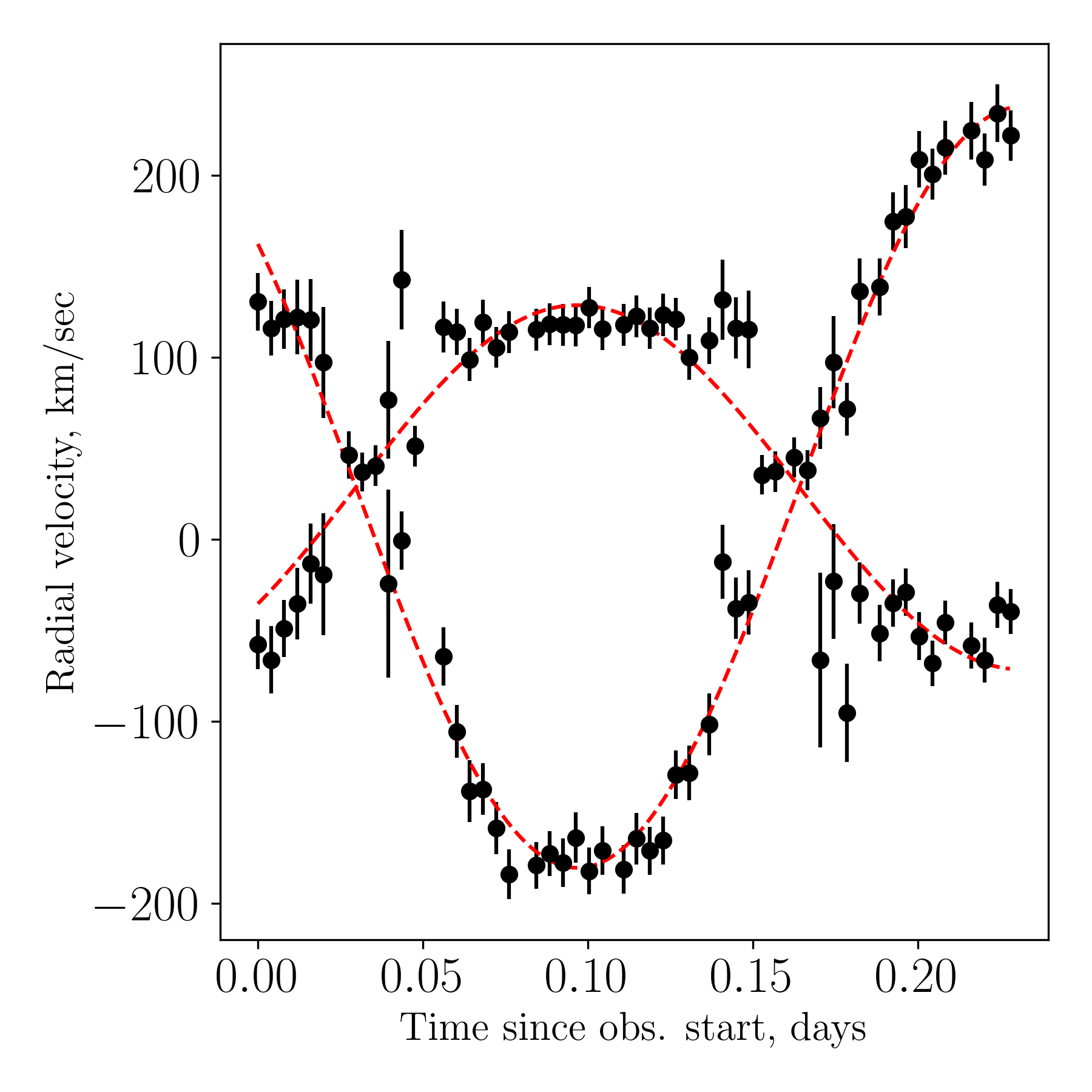}
        \caption{Absorption line centroids for the two components in the binary {\it Gaia} ID 1817286807102270336. Black points are the results of deblending the spectra into two spectroscopic components. The red dashed line is a circular orbital model with period fixed at the photometric value, $P=0.268858$ days. The fit yields a determination of the systemic velocity and of the amplitudes of the radial velocity variations of the components and therefore the masses of the components. The fit quality for {\it Gaia} ID 1817288284571020416, the other binary in the 2+2 system, is similar. The velocities shown are geocentric; the heliocentric correction on the night of the observations is $-23$ \kms \citep{Astropy2018}.}    
        \label{fig:double-lined}
    \end{figure}

    We directly confirm both 1817286807102270336 and 1817288284571020416 as double-lined spectroscopic binaries. Both binaries show high rotational broadening, $\sigma_{\rm rot}=90-100$ \kms, which significantly complicates deblending of the absorption features to obtain radial velocity curves for each. The rotational broadening can be constrained during the conjunction, when both stars in the binary are moving in the plane of the sky. After fixing rotational broadening to $\sigma_{\rm rot}$ and the period to the photometric period, it becomes possible to deblend the absorption lines using a model which consists of a systemic radial velocity for the binary, and time-varying flux contribution and time-varying radial velocity from each star. As a result, we can measure line centroids for each component (Figure \ref{fig:double-lined}). 

    With these results in hand, we further fit a circular orbit to each binary. The primary result is that the two binaries are moving with similar heliocentric radial velocities, $6\pm 5$ and $8 \pm 5$ \kms\ (errors combine in quadrature the error from the fit and the wavelength stability error from the analysis of the unrelated star). Therefore, this 2+2 system is a genuine comoving system despite a wide projected separation of 26,000 AU, and our spectroscopic observation directly validates our co-moving wide binary selection. Under the assumption that the orbit is exactly along the line of sight to the observer, the masses of the components are 0.55 and 0.27$M_{\odot}$ for {\it Gaia} ID 1817286807102270336, and 0.66 and 0.34$M_{\odot}$ for {\it Gaia} ID 1817288284571020416. These masses should be considered estimates because the spectroscopic observations did not cover one complete period for either binary and because of the difficulties deblending such strongly rotationally broadened spectra. 

\subsection{Enhancement of quadruple 2+2 systems}
\label{sec:enhancement}

    Our selection cannot capture any quadruple systems in which the two short-period components are separated by projected distances less than 0.2 arcsec, the spatial limit of {\it Gaia} source deblending. In addition, the selection of \texttt{phot\_bp\_rp\_excess\_factor}$<1.4$ for reliable {\it BP}$-${\it RP} colors also strongly reduce number of pairs at $<2$\,arcsec due to {\it BP} and {\it RP} photometry \citep{Riello2021}. Furthermore, we cannot identify quadruples with a high mass ratio between the two short-period components due to our color range selection. Finally, variability-based selection of the inner binaries is heavily biased toward shortest period objects and requires for the binary orbital plane to be close enough to the line of sight for the observer to see the eclipses. As a result, our sample of 2+2 systems is highly incomplete. Nonetheless, even with this limited sample we find an interesting result -- that the occurrence rate of 2+2 quadruples is significantly higher than that expected from random pairing up of field stars.

    The calculation of this `enhancement factor' starts with $F_{EB}$, the fraction of main-sequence sources that are EBs. The parent sample contains $N_{MS}=15,648,999$ target stars, and $N_{EB}=26,734$ of which are EBs with variability above 5\%. Therefore,  
     \begin{equation}
     \label{eq:f-eb}
         F_{EB} =\frac{N_{EB}}{N_{MS}} = \frac{26,734}{15,684,999} = (170\pm1)\times10^{-5},
     \end{equation}
     where the error is purely statistical and does not budget for any contamination of the EB sample.    
    
    The second step is to calculate $F_{2+2}$, the probability that a selected EB's companion is another EB. We calculate $F_{2+2}$, which is defined as follows: 
    \begin{equation}
    \label{eq:2+2}
        F_{2+2} = \frac{N_{2+2}}{N_{2+n}},
    \end{equation}
    where $N_{2+2}$ is the number of EBs in 2+2 systems, and $N_{2+n}$ is the number of EBs that have comoving companions in the parent sample regardless of whether the companion is eclipsing or not. We define $N_{2+2}$ and $N_{2+n}$ as the number of EBs, instead of the number of pairs. For example, with these definitions, there are $N_{2+2}/2$ 2+2 pairs, $N_{2+n}-(N_{2+2}/2)$ pairs that have at least one EB, and $N_{2+n}-N_{2+2}$ pairs that have exactly one EB. Also, one EB is counted as one in $N_{2+2}$ and $N_{2+n}$, even though there are actually two stars in one binary. 
    
    Because $N_{2+2}$ is always incremented by two (i.e. it is always an even number), we use a simulation to calculate the statistical uncertainty for $F_{2+2}$. We consider 0.5 million pairs (therefore 1 million stars) and randomly assign a star to be an EB with a certain probability. Then for the entire sample, we compute the ground-truth $F_{2+2, true}$ using Eq.~\ref{eq:2+2}. Then for the $i$-th experiment, we randomly draw a subset among the pairs so that its $N_{2+2,i}$ is on orders of 10 (similar to our observational results), and recompute $F_{2+2, i}$ for this subset. We repeat this experiment one thousand times and find that the standard deviation of $\{F_{2+2, i}\}$ is close to $\sqrt{2N_{2+2}}/N_{2+n}$. We thus adopt this as the uncertainty for $F_{2+2}$. 

    For the EBs selected to be $>$5\% variable, we have:
    \begin{equation}
        F_{2+2} = \frac{N_{2+2}}{N_{2+n}} = \frac{16}{1282} = (125\pm44)\times10^{-4}.
    \end{equation}
    In other words, for EBs that have comoving companions, about 1 out of 100 EBs has another EB companion. 

    Finally, we calculate the enhancement of the probability that an EB's companion is another EB compared to the probability that a main-sequence star is an EB:
    \begin{equation}
        \varepsilon_{2+2} = \frac{F_{2+2}}{F_{EB}} = \frac{0.0125}{0.00170} = 7.3\pm2.6.
        \label{eq:enhance}
    \end{equation}
    If the wide companions of EBs have the same EB fraction as the field stars, then $F_{2+2}=F_{EB}$ and $\varepsilon_{2+2}=1$. Thus, our results show that there is a factor of 7.3 enhancement in the pairing of two EBs compared to the random drawing of field stars from the parent sample.
    
    The definition of the enhancement factor in Eq.~\ref{eq:enhance} is robust against the incompleteness of EB selection and the contamination of chance alignments. If we are only recovering a fraction $f$ of the close binary population, the factor $f$ would appear in both $F_{2+2}$ and $F_{EB}$, canceling out in the calculation of enhancement factor. The chance-alignment pairs would make $F_{2+2}$ closer to $F_{EB}$, thus biasing the enhancement factor low. A lower chance-alignment contamination would only make the enhancement factor higher. Our investigation of phase-folded light curves also shows that our EB selection is efficient at selecting EBs with low contamination from non-eclipsing variables. Therefore, our measured enhancement factor is reliable over these possible systematics. 
    
    In \cite{Hwang2020short}, we found that there is a strong tail in variability distribution at $\gtrsim10$\% which is mainly due to short-period eclipsing binaries, and here we have found that the variability selection of $>5$\% still results in a very clean sample. If we try to repeat the analysis using $>3$\% selection, which is the instrumental noise level at the faintest magnitude in our sample, we recover 73 candidate 2+2 quadruple systems with a similar resulting enhancement factor of $\sim7$ with even smaller error compared to Eq.~\ref{eq:enhance}. However, we are not able to confirm the eclipsing binary nature from their light curves due to the quality of light curves in the current photometric surveys at this level of variability. The sample selected at $>$3\% variability may suffer a higher contamination from non-eclipsing variables, as evidenced by $>$3\%-selected variables being redder than $>$5\%-selected EBs. This suggests that $>$3\% variability may be due to stellar spots and flaring that are more common in redder stars. Therefore, in this paper, we focus on the results from the cleanest 5\% variability selection. 
    
\subsection{El-Badry et al. catalog}
\label{sec:elbadry}

    \citet{El-Badry2021} present a catalog of 1.3 million high-confidence wide binaries within 1\,kpc using {\it Gaia} EDR3. Their imposed criteria to identify candidate binary systems include (1) projected separations less than 1 pc; (2) parallaxes consistent within 3 or 6$\sigma$ for separations greater than or smaller than 4 arcsec, respectively; and (3) proper motions consistent with a Keplerian orbit. 
    
    We repeat the enhancement calculation of the double-close binary quadruples for this catalog. For both the primary and secondary pairs, we apply the same color range cut of \textit{BP$-$RP}=0.7-1.6\,mag, and the same criteria on {\it Gaia} photometry and astrometry described in Sec.~\ref{sec:gaia}. The fractional variability is computed following the same procedure in Sec.~\ref{sec:binaries}. \citet{El-Badry2021} estimate the probability of being a chance-alignment pair ($R$) for each wide binary, and we impose $R<0.1$ per their recommendation to select high-confidence wide binaries. We exclude white dwarfs by requiring the \texttt{binary\_type} parameter from \citet{El-Badry2021} be MS-MS. These selections result in 69,440 main-sequence wide binaries.
    
    Among these wide binaries, there are $N'_{2+n}=752$ EBs with fractional variability $>5$\% that have comoving companions, and $N'_{2+2}=10$ EBs that are in 2+2 systems (i.e. a total of five 2+2 systems). To distinguish the numbers from our comoving search, we use primes to indicate the numbers measured using the catalog from \citet{El-Badry2021}. All five 2+2 systems from \cite{El-Badry2021} are found in our comoving search. We have three additional 2+2 systems that are not in their catalog, and we mark these three systems in Table~\ref{tab:catalog5}. These three pairs all satisfy the relative velocity criteria used by \cite{El-Badry2021}, and they are likely removed by their other criteria. For example, \cite{El-Badry2021} exclude resolved triples, which may exclude the pair {\it Gaia} EDR3 1817286807102270336, 1817288284571020416 because it is a possible resolved 2+2+1 system. The other two pairs have small Galactic latitudes ($2.8$ and $-4.5$ degree), and may be excluded by \cite{El-Badry2021} due to their more strict source cleaning. 
    
    The probability of an eclipsing binary's companion being an eclipsing binary is $F'_{2+2}=N'_{2+2}/N'_{2+n}=10/752=(133\pm59)\times10^{-4}$. The enhancement of double-close binary quadruples calculated from this binary catalog is
    \begin{equation}
        \varepsilon'_{2+2} = \frac{F'_{2+2}}{F_{EB}} = \frac{0.0133}{0.00170} = 7.8\pm3.5,
    \end{equation}
    where $F_{EB}$ is from Eq.~\ref{eq:f-eb} because $F_{EB}$ does not depend on the comoving search. Therefore, despite the different comoving search approach in \cite{El-Badry2021}, we get a similar enhancement factor from their catalog.
    
\subsection{Zasche et al. catalog}

    A search for double-eclipsing binaries was performed in \citet{Zasche2019}. Their study mostly focuses on spatially unresolved photometric double-eclipsing binaries within the same light curve, but in addition they present a list of 5 candidate spatially resolved 2+2 systems. One of the systems, BV Dra \& BW Dra, is a known 2+2 system \citep{Batten1986} and is recovered by our search with {\it Gaia} EDR3 source IDs 1616589410627042560 \& 1616589410627042688. The other four systems were not recovered by our search due to the systems' parallaxes in the {\it Gaia} catalog falling below the limits of our search.

    %CoRoT 211659387 A: ra, dec: 286.00384112325, 3.50890380429
    %CoRoT 211659387 B: ra, dec: 286.00324465516, 3.50526385701

    We investigate the proper motion and parallax differences of these four systems using {\it Gaia} EDR3. Of the four unrecovered \citet{Zasche2019} 2+2 candidate systems, at least two are unlikely to be bound 2+2 quadruples. For CoRoT 211659387, the components have parallaxes $0.776\pm0.035$ and $0.334\pm0.082$\,mas, a 5-$\sigma$ difference. Their proper motion difference is $6.77\pm0.08$\masyr, corresponding to a relative velocity of $>41$\kms\ at a binary separation of $>10^4$\,AU, which is two orders of magnitude higher than a Keplerian orbit and is inconsistent with being a bound wide binary. For CzeV513 \& CzeV609 the parallaxes are $1.172\pm0.015$ and $0.899\pm0.018$ mas, inconsistent at 12-$\sigma$. With their proper motion difference of $2.86\pm0.03$\masyr, the relative velocity is 13\kms\ at $12$\,kAU if we adopt a system parallax of $1$\,mas, inconsistent with being a wide binary. In another 2+2 system, CzeV337 does not have parallax and proper motion available from Gaia EDR3, and thus we cannot confirm this pair as a genuine 2+2 system.
    
    The last object CRTS J065302.9+381408 \citep{Drake2014} may be a genuine bound wide binary. The components have consistent parallaxes ($0.291\pm0.026$ and $0.355\pm 0.051$ mas), with a proper motion difference of $0.18\pm0.05$\masyr, resulting a relative velocity of $2.94\pm0.90$\kms\ at $12$\,kAU, about 3-$\sigma$ higher than the expected Keplerian motion and would be selected by our demarcation line ($3.2$\kms\ at $12$\,kAU) if they were within our search volume. Therefore, out of the four 2+2 candidates identified in \citet{Zasche2019} that are not in our catalog, one is a wide binary candidate, one does not have Gaia astrometry available, and two are chance-alignment pairs.
    
\section{Discussion}
\label{sec:discussion}

\subsection{Dynamical evolution?}

    All eclipsing binaries in the 2+2 systems in Table~\ref{tab:catalog5} have orbital periods $<0.4$\,day, and they must have experienced orbital shrinkage on the main sequence because the orbital period at the time of formation must have been $\gtrsim30$\,yr, limited by the sizes of pre-stellar cores. Close binaries can shrink their orbit in the presence of outer companions through the Kozai-Lidov mechanism \citep{Kozai1962, Lidov1962,Naoz2016}. In this process, the inner binary evolves to a highly eccentric orbit, and at the close passage, the tidal friction can carry away the angular momentum, making the orbit smaller \citep{Kiseleva1998,Eggleton2001,Fabrycky2007,Naoz2014}. This process has been suggested for the formation of main-sequence contact binaries \citep{Borkovits2016}, although whether it is the main formation channel remains unclear \citep{Moe2018}. 
    
    In the 2+2 systems, one binary is the other binary's outer companion, and therefore both of the close binaries can undergo the Kozai-Lidov mechanism and form a double close-binary quadruple. This mutual Kozai-Lidov interaction can result in an enhancement of the 2+2 systems at short inner periods. However, we consider this scenario unlikely for most of the 2+2 systems reported in this paper. The main problem is that the 2+2 systems presented here have outer separations of $\gtrsim1,000$\,AU, which can only induce Kozai-Lidov oscillations when the inner binaries have separations $\gtrsim 5$\,AU limited by the relativistic pericenter precession \citep{Fabrycky2007}. The inner separations of the 2+2 systems reported here are $\sim0.01$\,AU, far too small for orbital shrinkage due to Kozai-Lidov mechanism with such a distant companion. 
    
    There are a few scenarios where this mutual Kozai-Lidov process can take place. The first possibility is that the two inner binaries initially had inner separations $\gtrsim5$\,AU so that the outer companions can trigger the Kozai-Lidov oscillations, and the inner binaries evolve to much smaller separations afterwards. The second possibility is that the outer separation was smaller before, and further evolved to the wide separation ($>1000$\,AU) that we observe here. This orbital widening can be due to the dynamical unfolding of compact multiples \citep{Reipurth2012}. We consider these two scenarios rather uncommon because they require very specific configurations of multiples, but a more detailed investigation is needed to firmly prove or disprove them. 

\subsection{Age and metallicity dependence}
\label{sec:dis-dependence}

    The enhanced close binary fraction in the wide companion of close binaries can be a result of the metallicity and age dependence of the close binary fraction. Since the component stars of binaries with separations $<1$ pc have (nearly) identical chemical abundances and ages \citep{Andrews2018,Andrews2019,Kamdar2019,Hawkins2020,Nelson2021}, if we find a close binary in a wide pair, then this wide pair is more likely to have a stellar age and/or metallicity that has a higher close binary fraction, thus resulting in a higher close binary fraction for its wide companion. In contrast, field stars have a wide range of metallicity and age, and some metallicity and age may have a lower close binary fraction, therefore lowering the average close binary fraction in the field star sample. 

    This explanation plays some role in the enhancement of 2+2 systems because the close binary fraction is indeed strongly dependent on metallicity and age. The close binary fraction is anti-correlated with metallicity \citep{Raghavan2010,Badenes2018,Moe2019, El-Badry2019a,Mazzola2020}. Furthermore, close binaries with orbital periods $<1$\,day strongly prefer stellar ages of a few Gyr \citep{Bilir2005,Li2007,Yildiz2014,Hwang2020short}. These dependencies can result in the enhanced fraction of 2+2 systems and imply that 2+2 systems are likely to have lower-than-average metallicities and have a narrower age distribution (centered at a few Gyr) than the field stars.

    However, the metallicity dependence can only explain a small fraction of the 2+2 enhancement. The metallicity of our eclipsing binaries is unknown, but it is likely lower than the mean metallicity of field stars due to the anti-correlation between the metallicity and the close binary fraction, but it cannot be too low because there is a deficit of short-period binaries with tangential velocities $>100$\kms, which are mostly thick-disk and halo stars \citep{Hwang2020short}. By cross-matching the LAMOST survey \citep{Deng2012,Zhao2012} with {\it Gaia}, we find that the mean metallicity for the field stars is [Fe/H]$=-0.15$, and the mean metallicity at tangential velocities of $100$\kms\ is [Fe/H]$=-0.3$. If the mean metallicity of the eclipsing binaries is lower than the field stars by $\Delta$[Fe/H]$=0.15$, then the close binary fraction at the metallicity of the eclipsing binaries is only a factor of 1.12 higher than the fraction at the metallicity of the field stars \citep{Moe2019}. This 1.12 enhancement from the metallicity dependence is much smaller than the measured enhancement of $\sim7$, meaning that some other effects need to account for a factor of $7/1.12\sim6$ enhancement. 
    
    Here we investigate the level of enhancement resulting from the age dependence of short-period binaries. Specifically, we use the {\it Gaia} mock catalog \citep{Rybizki2018} which is generated from the Besan\c{c}on Galactic model \citep{Robin2003} so that stellar ages are available. Then we consider a simple lifetime model for short-period eclipsing binaries, where all short-period binaries are formed at an age of $t_0$, and merge at an age of $t_1$. \cite{Hwang2020short} use the Galactic kinematics to constrain $t_0$ and $t_1$, showing that $t_0=0.6$-3\,Gyr and $t_1<10$\,Gyr, with several combinations of $t_0$ and $t_1$ that can describe the observational results. In addition to $t_0$ and $t_1$, another free parameter is the intrinsic eclipsing binary fraction (IEBF), which is the fraction of stars that would become an EB during their lifetime. Under the assumption that all EBs have the same formation time and merger time, IEBF is equal to $F_{2+2}$. Then we can compute the enhancement factor $\varepsilon_{2+2}$ for all accepted $t_0$ and $t_1$ combinations.
    
    We find that only when $t_1 - t_0$ is small can we obtain a high enhancement factor in this model. This is because a small $t_1 - t_0$ -- i.e., short net life-time of EBs -- results in a lower $F_{EB}$ in the field stars while also requiring a higher IEBF to be consistent with the observed eclipsing binary fraction. Specifically, only ($t_0$, $t_1$)$= (3, 5)$ and $(3, 6)$\,Gyr have an enhancement $>3$, with $\varepsilon_{2+2}=5.2$ and 3.7 respectively. 
    
    In reality, EBs are not formed and merged at the same time. The formation time depends on the initial condition of the binary (e.g. initial separation) and on its physical properties like the mass. Then for a small $t_1 - t_0$, for example with ($t_0$, $t_1$)$= (3, 5)$\,Gyr, a small formation time difference of 1\,Gyr would strongly reduce the enhancement by a factor of 2, changing $\varepsilon_{2+2}$ to  $5.2/2=2.6$. Taking the formation and merger time distribution into account, the combinations of $(t_0,t_1)$ with a small $t_1 - t_0$ would have reduced the enhancement factor to $\sim2$, while combinations of $(t_0,t_1)$ with larger $t_1 - t_0$ are less sensitive to the different formation times and may still have an enhancement between 1 and 3. Therefore, the age dependence results in a enhancement by a factor less than 3.
    
    In conclusion, the metallicity dependence of close binary fraction may only account for an enhancement of at most of a factor of 1.12, and the age dependence for an enhancement of at most a factor of 3. Their combination therefore can only explain an enhancement of no more than 3.36, and this is still likely an over-estimate because of the finite formation and merger time distribution. While this factor may be marginally consistent with our enhancement measurement due to its sizable measurement uncertainty, we discuss other possibilities that may contribute to the enhancement. 

\subsection{Special formation channel for EBs in wide pairs?}

    It is thought that close and wide binaries are formed in different ways. Binaries with separations $\lesssim10$\,AU are mostly formed from the disk fragmentation \citep{Kratter2006,Tanaka2014}. In contrast, wide binaries with separations $>100$\,AU can be formed from turbulent core fragmentation \citep{Padoan2002, Fisher2004, Offner2010}, dynamical unfolding of compact triples \citep{Reipurth2012}, the dissolution of star clusters \citep{Kouwenhoven2010, Moeckel2011}, and the random pairing of pre-stellar cores \citep{Tokovinin2017}. Different binary formation mechanisms dominating at close and wide separations are supported by the different metallicity dependences of the close and wide binary fractions. Specifically, the close binary fraction (separations $<100$\,AU) is anti-correlated with metallicity \citep{Raghavan2010,Badenes2018,Moe2019, El-Badry2019a,Mazzola2020}, and the wide binary fraction (separations between $10^3$ and $10^4$\,AU) has a non-monotonic dependence on the metallicity \citep{Hwang2021a}, peaking around the solar metallicity and decreasing toward both low and high metallicity ends. Finally, close and wide binaries have completely different eccentricity distributions \citep{Hwang2021b}. Therefore, in terms of the overall population, close and wide binaries are likely formed differently and should be uncorrelated.

    In this paper, we show that for EBs with orbital periods $<1$\,day, their occurrence rate is positively correlated with their wide companions, resulting in the enhanced 2+2 systems. These systems constitute a small subset of the binary population and thus they do not contradict the different metallicity dependence of the overall close and wide binary fraction. In Eq.~\ref{eq:enhance}, we find that the occurrence rate of 2+2 systems is 7 times higher than random pairing of field stars. Here we investigate the occurrence rate of EBs in wide pairs, $F_{2+n}$. Based on our results, $F_{2+n} = N_{2+n}/N_{wide}= 1282/ (262,490)=0.49\pm0.01$\%, where $N_{wide}$ is the number of sources (EB or otherwise) in wide comoving pairs. Then the enhancement factor $\varepsilon_{2+n}$, defined as the ratio between the occurrence rate of EBs in wide pairs to that in the field stars, is
    \begin{equation}
        \varepsilon_{2+n}= \frac{F_{2+n}}{F_{EB}} = \frac{0.0049}{0.0017}=2.88\pm0.08,
    \end{equation}
    showing that the occurrence rate of EBs among wide pairs is a factor of $\sim3$ higher than that in the field stars. $\varepsilon_{2+n}$ is equivalent to the ratio between the wide companion fraction of EBs ($N_{2+n}/N_{EB}$) to that of the field stars ($N_{wide}/N_{MS}$), where \cite{Hwang2020wide} measure to be $3.1\pm0.5$ with a slightly different sample selection, consistent with the value reported here. 
    
    Therefore, a factor of 7 enhancement of the 2+2 systems ($\epsilon_{2+2}$ in Eq.~\ref{eq:enhance}) can be explained by (1) a factor of 3 ($\epsilon_{2+n}$) from the higher occurrence rate of close binaries in wide pairs, and (2) a factor of $\sim2$ from the age and metallicity dependence of the binary fraction. It remains uncertain what exactly makes EBs three times more likely to form in wide pairs than in the field stars. One possibility is that, for $2+n$ and $2+2$ systems, the formation of close binaries and wide companions is correlated \citep{Hwang2020wide}. For example, inner binaries and outer companions can be formed together from the dynamical unfolding of compact multiples \citep{Reipurth2012}, or alternatively, the presence of wide companions may facilitate the disk fragmentation and enhance close binary formation \citep{Tokovinin2021}.

    Our measurements suggest that a special formation channel for EBs in wide pairs and 2+2 systems needs to account a factor of $\sim3$ enhancement in addition to the effects of the metallicity and age to fully explain the observation. \cite{Tokovinin2014} also find that the observed 2+2 systems are $\sim2.5$ times more frequent than his simulation that assumes uncorrelated formation. Our 2+2 systems are extreme, with inner separations of $\sim0.01$\,AU and outer separations from $\sim10^3$\,AU up to $10^4$\,AU, providing an important clue that this special formation channel can operate for a wide range of separation scales. 

\subsection{How long can contact binaries stay in contact?}

    Most of the eclipsing binaries reported here have orbital periods $<0.5$\,day. At this short period end, most ($\sim80$\%) of the main-sequence eclipsing binaries are contact binaries \citep{Paczynski2006}. Indeed, all our light curves show light curves that are characteristic of contact binaries, although we do not attempt to conduct light curve fitting to distinguish contact binaries and semi-detached binaries.
    
    How long a contact binary can stay in contact remains an unsolved question. A wide range of timescales have been suggested in the literature, from 0.1-1 Gyr \citep{Guinan1988}, 1.61 Gyr \citep{Bilir2005}, to 7.2 Gyr \citep{Li2005}. The fact that we can find a sizable number of 2+2 systems suggests that contact binaries can stay in contact for a significant fraction of their age. Otherwise, if contact binaries could only stay in contact for a duration much shorter than their age, chances would be too low to capture two binaries in the contact configuration at the same time. With a mean age of a few Gyr for contact binaries based on their kinematics \citep{Bilir2005, Hwang2020short}, our result suggests that the timescale that a contact binary can stay in contact is also about a few Gyr. The upper limit on this timescale is about 10\,Gyr because in the kinematically old population (thick disk and halo) the contact binary fraction is significantly lower than that in the thin disk of the Galaxy, and therefore contact binaries must merge and disappear from the population \citep{Hwang2020short}. 

\section{Conclusions}
\label{sec:conclusions}

    In this paper, we present a new sample of 8 hierarchical 2+2 quadruples from {\it Gaia} EDR3, with 7 of them being newly discovered. Our selection is sensitive to quadruples with outer separations $\gtrsim 1000$ AU, while the inner binaries are identified via their variability due to mutual eclipses, and therefore typically have short (near-contact) periods of half a day or less. We find that the wide companions of eclipsing binaries are $7.3\pm2.6$ more likely to be another eclipsing binary compared to the field stars.
    
    We then analyze the light curves of the 7 newly discovered 2+2 systems using {\it WISE}, ZTF, and ASAS-SN. We successfully recover the orbital periods for all but one EB (Fig.~\ref{fig:lightcurves1}). Their orbital parameters are presented in Table~\ref{tab:catalog5}.  
    
    In our 2+2 systems, the outer and inner separations are different by 5-6 orders of magnitude. As a result, we do not expect any relationship between the inner periods as the binaries are too far apart for each to be affected by the internal structure of the other. In particular, there is no evidence for resonances that are seen in 2+2 quadruples with much closer outer separations and identified from their blended light curves \citep{Zasche2019, Tremaine2020}, and for the systems where both inner periods are available, we find that the period ratios are drawn from a distribution consistent with random pairings of available periods. The period ratios of inner binaries are not consistent with a $\propto \left(P_{>}/P_{<}\right)^{-2}$ distribution expected if the periods are drawn from a uniform distribution \citep{Tremaine2020}, with the Kolmogorov-Smirnov probability of the null hypothesis of $0.03\%$, but this result can be explained by our selection bias toward short-period systems and therefore the period distribution is not uniform.
    
    Given that we expect little to no dynamical interactions between the pairs, it is with some surprise that we find that the widely separated, co-moving 2+2 quadruples are about $7.3\pm2.6$ more common than would be expected from random pairings of field stars (singles and binaries). An enhancement of 2+2 quadruples is also pointed out in a volume-limited sample by \citet{Tokovinin2014}, who suggests that these systems are formed by a special process. Similarly, we find that the metallicity and age dependence of the close binary fraction can only account for at most half of the observed enhancement. Therefore, our results suggest that a special formation channel is needed to further enhance the close binary formation in wide pairs, thus enhancing the 2+2 populations, by another factor of 3. Our results also imply that contact binaries are long-lived, placing an important constraint on the open question on the stability of contact binaries. 

\section*{Acknowledgments} 
    This work has made use of data from the European Space Agency (ESA) mission
    {\it Gaia} (\url{https://www.cosmos.esa.int/gaia}), processed by the {\it Gaia} Data Processing and Analysis Consortium (DPAC, \url{https://www.cosmos.esa.int/web/gaia/dpac/consortium}). Funding for the DPAC has been provided by national institutions, in particular the institutions participating in the {\it Gaia} Multilateral Agreement.
    
    This publication makes use of data products from the Wide-field Infrared Survey Explorer, which is a joint project of the University of California, Los Angeles, and the Jet Propulsion Laboratory/California Institute of Technology, funded by the National Aeronautics and Space Administration.
    
    This publication is based in part on observations obtained with the Samuel Oschin 48-inch Telescope at the Palomar Observatory as part of the Zwicky Transient Facility project. ZTF is supported by the National Science Foundation under Grant No. AST-1440341 and a collaboration including Caltech, IPAC, the Weizmann Institute for Science, the Oskar Klein Center at Stockholm University, the University of Maryland, the University of Washington, Deutsches Elektronen-Synchrotron and Humboldt University, Los Alamos National Laboratories, the TANGO Consortium of Taiwan, the University of Wisconsin at Milwaukee, and Lawrence Berkeley National Laboratories. Operations are conducted by COO, IPAC, and UW.
    
    This publication is based in part on observations obtained with the Apache Point Observatory
	3.5-meter telescope, which is owned and operated by the Astrophysical
	Research Consortium.
	
    This publication is based in part on data from ASAS-SN, which is supported by the Gordon and Betty Moore Foundation through grant GBMF5490 to the Ohio State University, and NSF grants AST-1515927 and AST-1908570. Development of ASAS-SN has been supported by NSF grant AST-0908816, the Mt. Cuba Astronomical Foundation, the Center for Cosmology and AstroParticle Physics at the Ohio State University, the Chinese Academy of Sciences South America Center for Astronomy (CASSACA), the Villum Foundation, and George Skestos.
    
     The authors thank the referee J.Chaname for the positive report and constructive comments that helped improve the manuscript. H.-C.H. thanks A.Tokovinin for discussions and comments on the manuscript. G.B.F., H.-C.H., and N.L.Z. were supported in part by Space@Hopkins and NASA ADAP grant 80NSSC19K0581. H.-C.H. acknowledges the support of the Infosys Membership at the Institute for Advanced Study.

\section{Data Availability} 
    \textit{The data underlying this article are available in the article and in its online supplementary material.}

\bibliography{references}{}
\bibliographystyle{aasjournal}

\appendix
\section{Gaia Query} \label{query}
    This is the Gaia query described in Section \ref{sec:gaia}:
    
    \begin{verbatim}
    SELECT 
        gaia.*
    FROM gaiaedr3.gaia_source AS gaia
    WHERE
        (gaia.astrometric_params_solved = 31 OR gaia.astrometric_params_solved = 95) AND
        gaia.parallax_over_error > 5 AND
        gaia.phot_g_mean_flux_over_error>50 AND
        gaia.phot_bp_mean_flux_over_error>20 AND
        gaia.phot_rp_mean_flux_over_error>20 AND
        gaia.visibility_periods_used>=11 AND
        gaia.ruwe < 2 AND
        gaia.bp_rp> 0.7 AND
        gaia.bp_rp< 1.6 AND
        gaia.parallax > 1 AND
        gaia.phot_bp_rp_excess_factor < 1.4
    \end{verbatim}

\end{document}